\documentclass{ws-ijmpa}
\usepackage[super,compress]{cite}
\usepackage{graphicx}
\usepackage{color,amsmath,amsfonts,multirow}
\usepackage{epsfig}
\usepackage{rotating}
\usepackage{array}
\usepackage{graphicx}
\usepackage{multirow}
\usepackage{epstopdf}
\usepackage{amssymb}
\usepackage{latexsym}
\usepackage{comment}
\usepackage{bm}
\usepackage{mathtools}
\usepackage{xspace}
\usepackage{subfigure}
\def\be{\begin{eqnarray}}
\def\ee{\end{eqnarray}}
\newcommand{\fig}{\begin{figure}}
\newcommand{\ef}{\end{figure}}

\newcommand{\nua}[1]{\ensuremath{\rlap{\kern-2.5pt\ensuremath{\overset{\scriptscriptstyle(-)}{\phantom{\nu}}}}{\ensuremath{{\nu}_{#1}}}}\xspace}
\newcommand{\lcdm}{$\Lambda$CDM\xspace}
\newcommand{\Neff}{\ensuremath{N_{\eff}}\xspace}
\newcommand{\eff}{\ensuremath{\mathrm{eff}}\xspace}
\newcommand{\thm}{\ensuremath{\mathrm{th}}\xspace}
\newcommand{\DNeff}{\ensuremath{\Delta N_{\eff}}\xspace}
\newcommand{\meff}[1]{\ensuremath{m^{\eff}_{#1}}\xspace}

\newcommand{\nbd}{$0\nu\beta\beta$}
\begin{document}
\markboth{Sin Kyu Kang}{Roles of sterile neutrinos in particle physics and cosmology}

%
\catchline{}{}{}{}{}
%

\title{Roles of sterile neutrinos in particle physics and cosmology 
}

\author{Sin Kyu Kang
}

\address{School of Liberal Arts, Seoul Tech, Gongreung-ro 232\\
Seoul, 01811, Republic of Korea
\\
skkang@snut.ac.kr}

\maketitle

\begin{history}
\received{Day Month Year}
\revised{Day Month Year}
\end{history}

\begin{abstract}

The impacts of the light sterile neutrino hypothesis in  particle physics and cosmology are reviewed.
The observed short baseline neutrino anomalies challenging  the standard explanation of neutrino oscillations within the framework of three active neutrinos are addressed.
It is shown that they can be interpreted as the experimental hints pointing towards the existence of sterile neutrino at the eV scale.
While the electron neutrino appearance and disappearance data are in favor of such a sterile neutrino,
the muon disappearance data disfavor it, which gives rise to a strong appearance-disappearacne tension.
After a brief review on the cosmological effects of light sterile neutrinos,  proposed signatures of light sterile neutrinos in the existing cosmological data are discussed.
The keV-scale sterile neutrinos as possible dark matter candidates are also discussed by reviewing  different mechanisms of how they can be produced in the early Universe and how their properties can be constrained by several cosmological observations.
We give an overview of the possibility that  keV-scale sterile neutrino can be a good DM candidate and  play a key role in achieving low scale leptogenesis simultaneously by introducing a model where an extra light sterile neutrino is added on top of type I seesaw model.

\end{abstract}

\ccode{PACS numbers:14.60.St,98.80.−k,95.35.+d}


\section{Introduction}	
The standard model (SM) of elementary particles is the unified theory of three fundamental interactions,
the electromagnetic, weak and strong interactions in the universe as well as classifying all known elementary particles.
Developed in the early 1970s, it has successfully explained almost experimental results and precisely predicted a wide variety of phenomena, which make it a well-tested physics theory.
Yet,  it has been increasingly clear that apart from being theoretically unsatisfactory the SM fails to explain a number of the oberved phenomena in particle physics, astrophysics and cosmology.

Neutrinos, the most mysterious and fascinating of all elementary particles, continue to puzzle physicists. 
It is common belief that
they have a very important role in both of the microscopic view of particle physics and the macroscopic view of evolution of the universe. 
The remarkable development over the past two decades  has been the discovery of the flavor oscillations in neutrinos\cite{atm,solar,re1,re2,re3},  which has established that neutrinos are massive and mix.

The existence of three active neutrinos has been confirmed by the precision measurement of the partial decay width $\Gamma(Z^0\rightarrow \nu \bar{\nu})$ leading us to determine the number of light neutrino species, $N_{\nu}=\frac{\Gamma_{inv}}{\Gamma(Z^0\rightarrow \nu \bar{\nu})}=2.9840\pm0.0082$~\cite{ALEPH:2005ab} with $\Gamma_{inv}$ being the invisible decay width of $Z^0$.
The result of $N_{\nu}$ gives the number of light particles that have the standard properties of neutrinos with respect to the weak interactions.
In the framework of three active neutrinos, neutrino oscillation can be described by 6 independent parameters
\cite{review,pdg}, in which
5 of them, three mixing angles and two mass squared differences, have been measured by
the experiments aimed to detect the solar, atmospheric, reactor and accelerator neutrinos\cite{atm,solar,re1,re2,re3} . 
The remaining parameter in the mixing matrix, CP violating phase, becomes the main target of next generation of neutrino experiments \cite{CPV}.
Although neutrinos have turned out to be massive particle, their individual masses have not been measured at all.
The hierarchy of their masses is also one of big mysteries, as the ordering of their values has profound implication in particle physics and cosmology \cite{Vagnozzi:2017ovm,mass-ordering2}.

On the other hand, there are other issues that keep neutrino physicists puzzled, which
are known to be {\it neutrino anomalies} that are unexpected results in several experimental studies
as will be discussed in detail later.
They challenge the standard explanation of neutrino oscillations which have been used to determine
three mixing angles and two mass squared differences in the framework of three active neutrinos.
There have been a large number of theoretical attempts to interpret these as signs of new physics beyond the standard paradigm. 
The most intriguing explanation of the observed anomalies involves a forth type of massive neutrino separated from the three others by a new mass squared difference larger than 0.1 $\rm eV^2$ \cite{Giunti:2019aiy}.
It is called {\it sterile neutrino} hypothetical particle \cite{Pontecorvo:1967fh}  that interacts in the cosmological context only via gravitation and through mixing the other neutrinos avoiding any of the fundamental interactions of the SM.

The study of sterile neutrinos presents an interesting avenue into new physics beyond the SM \cite{Volkas:2001zb,Mohapatra:2006gs,Boyarsky:2009ix,Giunti:2015wnd,Gariazzo:2015rra,Adhikari:2016bei,Abazajian:2017tcc,Lattanzi:2017ubx,Boyarsky:2018tvu,Giunti:2019aiy}. 
Apart from the experimental motivation mentioned above, there are a few good reasons to postulate the existence of sterile neutrinos: to explain the origin of tiny neutrino masses\cite{seesaw}, dark matter candidate \cite{Dodelson:1993je,Boyarsky:2018tvu}and role in baryogenesis \cite{leptogenesis}. 
%
Sterile neutrinos usually refer to neutrinos with right-handed chirality, which may be added to the SM
to endow neutrinos with mass in a simple way \cite{seesaw}.
%
The masses of the right-handed neutrinos are unknown and could have any value between $10^{15}$ GeV and sub eV.  
But, sterile neutrinos can have masses around the electroweak scale and their Yukawa couplings with the active neutrinos and the Higgs boson are unsuppressed when they are subject to a protective symmetry \cite{Antusch:2016ejd,Mohapatra:1986bd}.
In such a case, sterile neutrinos can affect some phenomena occurred from high energy collisions and thus
would be testable at high energy colliders \cite{Han:2006ip,Atre:2009rg,Dev:2013wba,Alva:2014gxa,Antusch:2016ejd}.
Cosmology has an important role to play here, since the constraints obtained from cosmological observations are quite complementary to those coming from particle experiments \cite{Adhikari:2016bei,Abazajian:2017tcc,Boyarsky:2018tvu}.
The existence of sterile neutrinos may affect the evolution of the Universe.
They may serve as a good dark matter (DM) candidate and play an important role in making the Universe matter-dominant \cite{Adhikari:2016bei,Abazajian:2017tcc,Boyarsky:2018tvu}.

%
%
The aim of this review is to show how the possibe existence of light sterile neutrinos may affect phenomena
in particle physics and cosmology.
For the light sterile neutrinos, we consider two intriguing cases depending on the magnitude of their mass,  eV  and keV scales.
The study of eV scale sterile neutrino has been popular due to some anomalies explained in terms of neutrino oscillations including sterile neutrino.
We review on how those neutrino anomalies can be explained in terms of neutrino oscillations in the framework of the so-called $3+1$ model,
and how the existence of light sterile neutrino can affect cosmology such as Big Bang Nucleosynthesis (BBN), Cosmic Microwave Background (CMB) and Large Scale Structure (LSS) snd so on.
The sterile neutrinos with a mass of  keV scale  have been attracted much attention due to some cosmological observations and possible dark matter candidate.
They can also play some roles in cosmology such as baryogenesis through leptogenesis.
In particular, we give an overview of the possibility that  keV-scale sterile neutrino can be a good DM candidate and  play a key role in achieving low scale leptogenesis simultaneously by introducing a complete model .

This article is organized as follows. In section 2, we study the implications of the existence of light sterile neutrino by reviewing the current status of short base-line (SBL) experiments which lead to the unexpected
results challenging the standard explanation of neutrino oscillation in the framework of three active neutrinos.
We discuss whether a sterile neutrino with a mass of eV scale added on top of three active neutrinos can play  a crucial role in reconciling those anomalous experimental results in terms of neutrino oscillations.
In section 3, we study some roles of light sterile neutrinos in cosmology and discuss how their properties can be constrained by several cosmological observations. We introduced a model where tiny neutrino masses, low scale leptogenesis as well as sterile neutrino DM with a mass of keV scale are simultaneously accommodated.
Finally, we summarize the roles of sterile neutrinos in particle physics and cosmology discussed in this review and give a short perspective of future study of light sterile neutrinos.

\section{Light Sterile Neutrino in Particle Physics}
Although eV scale sterile neutrino is not theoretically well motivated, an indication of possible existence of
sterile neutrinos has arised from some anomalous experimental observations
unexplained in terms of neutrino oscillations in the framework of three active neutrinos.
Those anomalies have emerged
 in several SBL neutrino experiments with the ratio of the oscillation distance over the
neutrino energy, $L/E\sim 1 m/{\rm MeV}$, which are sensitive to neutrino oscillations involving mass squared differences $\Delta m^2 \sim 1~\mbox{eV}^2$.
We review the current status on the neutrino anomalies and how they can explained in terms of neutrino oscillations including sterile neutrinos with a mass of eV scale.
%
%
\subsection{The LSND and MiniBooNE Anomalies}
An experiment at Los Alamos National Laboratory (LSND) \cite{Athanassopoulos:1996jb} created a large number of low-energy muon antineutrinos, $\bar{\nu}_{\mu}$, from the production of positive pions.
 While in the chain of reactions  muon neutrinos and antineutrinos are produced in equal numbers, if the positive muons are stopped before they decay, the resulting antineutrinos populate the energy spectrum between 20 and 60 MeV where the contamination from their antiparticles can be kept smaller than 1 per mille. LSND could thus search for the appearance of electron antineutrinos, $\bar{\nu}_e$,  traveled about $30 m$  in a large tank containing 167 tons of liquid scintillator, where the inverse beta decay reaction, $\bar{\nu}_e p\rightarrow e^+ n$,  takes place with very small backgrounds. 
%
In five years of data taking, from 1993 to 1998, LSND collected  $90\pm23$ $\bar{\nu}_e$ appearance events at the $3.8 \sigma$ level. 
The anomalous result observed at the LSND experiment \cite{Aguilar:2001ty} was the first indication of the possible existence of  light sterile neutrino.
It has been shown that the result can be interpreted in terms of a new mixing angle $\theta$ and a new mass-squared difference $\Delta m^2\sim 1 \mbox{eV}^2$. 
But, the independent experiment KARMEN \cite{Armbruster:2002mp} designed similar to LSND observed no such an anomalous  signal, even though it was not able to rule out completely the regions of the parameters to interpret  the LSND result in terms of oscillations.

The MiniBooNE experiment at Fermilab \cite{AguilarArevalo:2007it} was designed to test the LSND results.
In contrast to LSND, MiniBooNE made use of a conventional neutrino beam  with neutrino energy of 600 (400) MeV and the same $L/E_{\nu}$ as in LSND 
produced from mesons decaying in flight which generate at Fermilab.
The MiniBooNE experiment has searched for $\nu_{\mu}\rightarrow \nu_e$ (or $\bar{\nu}_{\mu}\rightarrow \bar{\nu}_e$) oscillations
by measuring the rate of $\nu_e n \rightarrow e^- p$ (or $\bar{\nu}_e p \rightarrow e^+ n$) events and testing the consistency of the measured rate with the estimated background one.
The first result reported in 2007 came out from the search of neutrino mode and showed no excess of events
with $E_{\nu}>475$ MeV, which was somewhat inconsistent with the LSND result for anitneutrinos.
This was followed by the result released in 2009 \cite{AguilarArevalo:2008rc} showing an unexplained excess of $\nu_e$ events
below 475 MeV.
Recently, MiniBooNE reported the observation of a total electron neutrino event excess in both neutrino and antineutrino modes of $460.5\pm 95.8$ events quantified at the level of 4.8 $\sigma$ from the background-only hypothesis \cite{Aguilar-Arevalo:2018gpe}.
The observed excess is in agreement with the LSND result, and provides a good fit to a $\Delta m^2$ of order  $1 ~{\rm ev}^2$ in a two-neutrino oscillation framework. 
Combining the two results leads to a global significance exceeding $6 \sigma$~ \cite{Aguilar-Arevalo:2018gpe} .

An independent test of the LSND and MiniBooNE anomalies has been recently carried out at the long-baseline experiments ICARUS \cite{Antonello:2012pq, Farnese:2015kfa} and OPERA\cite{Agafonova:2013xsk}. 
In these experiments, due to the high energy of the beam, $< E > \sim 17$ GeV, the 3-flavor effects induced by non-zero $\theta_{13}$ are negligible, which makes them sensitive to sterile neutrino oscillations.
But, the current sensitivities of  ICARUS and OPERA are not enough to rule out the mass-mixing region preferred by the LSND and MiniBooNE, and they can only restrict the allowed region to values of $\sin^2 2\theta$  less than a few $\times 10^{-2}$.

\subsection{Reactor Antineutrino and Gallium Anomalies}
A second class of anomalies was found in reactor-based experiments searching for the disappearance of $\bar{\nu}_e$ from the reactor flux  \cite{reactorantineutrinoanomaly}, as well as in the measurement of solar neutrino by the Gallium experiments, GALLEX\cite{Kaether:2010ag} and SAGE\cite{Abdurashitov:2005tb}. 
A deficit of $\bar{\nu}_e$ fluxes observed at the distances $10-100~m$ between reactors and detectors in several SBL neutrino experiments in comparison with that expected from new calculations \cite{Mueller:2011nm,Huber:2011wv} is known now as the {\it reactor antineutrino anomaly}. 
The new calculation based on more sophisticated {\it ab-initio} method improved  $\bar{\nu}_e$ fluxes that are about $5\%$ larger than the previous one\cite{previous}.
Recently, new reactor antineutrino spectra from the four main fission isotopes $^{235}{\rm U}, ^{239}{\rm Pu}, ^{241}{\rm Pu}$, and $^{238}{\rm U}$ have been provided, which increase the mean flux by about 3 $\%$ \cite{Mueller:2011nm}. To a good approximation, this reevaluation applies to all reactor neutrino experiments giving
new impetus to the study of light sterile neutrinos.
The ratio of measured to predicted antineutrino flux based on the new calculation showed a deficit of $6\%$\cite{reactorantineutrinoanomaly}.
%
\begin{figure}[t]
\begin{center}
\includegraphics*[width=\textwidth]{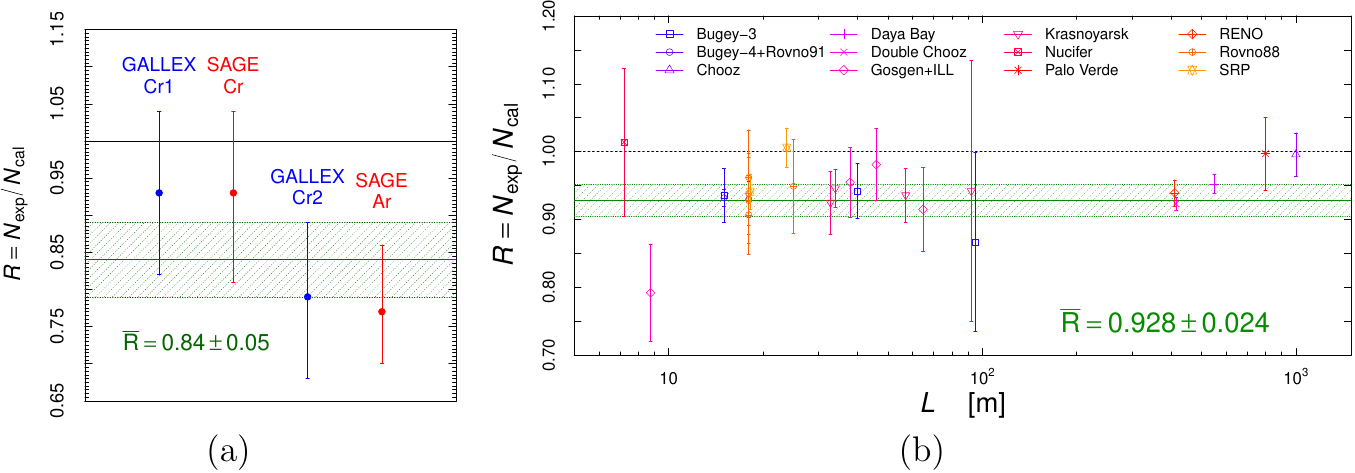}
\end{center}
\caption{\label{fig:galrea}
Plots for the Gallium neutrino (a)
and reactor antineutrino (b)
anomalies \cite{Giunti:2019aiy}.
}
\end{figure}
%

In Figure~\ref{fig:galrea},  the ratios $R$ of the measured ($N_{\text{exp}}$) and calculated ($N_{\text{cal}}$)
number of $\bar{\nu}_e$ events are plotted observed in the various experiments :
Bugey-3 \cite{Declais:1995su},
Bugey-4 \cite{Declais:1994ma},
ROVNO91 \cite{Kuvshinnikov:1990ry},
Chooz \cite{Apollonio:2002gd},
Gosgen \cite{Zacek:1986cu},
ILL \cite{Kwon:1981ua,Hoummada:1995zz},
Krasnoyarsk \cite{Vidyakin:1994ut},
Nucifer \cite{Boireau:2015dda},
Rovno88 \cite{Afonin:1988gx},
SRP \cite{Greenwood:1996pb},
Palo Verde \cite{Boehm:2001ik},
Daya Bay \cite{An:2017osx,Adey:2018qct}, 
RENO \cite{RENO:2018pwo}, and
Double Chooz \cite{Abe:2014bwa,Bezerra-NOW2018}.
As presented in the figure, the global world average of the ratio is  $\overline{R} = 0.928 \pm 0.024$,
which corresponds to a $3.0\sigma$ deficit \cite{Giunti:2019aiy}.
Given the ranges of reactor neutrino energies and distances between sources and detectors,
the explanation of 
the deficit in terms of neutrino oscillations
requires $\Delta{m}^2_{\text{SBL}} \gtrsim 0.5 \, \text{eV}^2$, which is consistent
with the LSND and MiniBooNE signals \cite{Giunti:2019aiy}.
It is worthwhile to note that the reactor experiments could only see the averaged effect of
the $L/E_{\nu}$ modulation due to lack of the energy resolution, so
the deficit might be due to flux mismodellings.

The explanation of the reactor antineutrino anomaly in terms of neutrino oscillations can be supported by 
some experimental results that are free from flux calculations.
NEOS \cite{Ko:2016owz} and DANSS \cite{Alekseev:2018efk} experiments performed measurements of $\bar{\nu}_e$ fluxes so as to be free from flux calculations and reported some wiggles in their spectra that can be fitted well by neutrino oscillation
although the results do not have enough statistical significance yet.
The coincidence of the NEOS \cite{Ko:2016owz}and DANSS \cite{Alekseev:2018efk} best fits at $\sin^2 2 \theta\simeq 0.04-0.05$ and
$\Delta m^2\simeq 1.3 - 1.4~{\rm eV}^2$ shows an indication  in favor of SBL active-sterile neutrino oscillations \cite{Giunti:2019aiy}.

In addition to the flux discrepancy, SBL experiments such as Daya Bay\cite{An:2015nua}, RENO\cite{RENO:2015ksa}, Double Chooz \cite{Abe:2014bwa} and NEOS \cite{Ko:2016owz}
observed statistically significant excess over prediction in the $\bar{\nu}_e$ spectrum around 5 MeV
at the  $\sim 4 \sigma$ level. 
The bump structure appears to be similar at the near and far detectors and is correlated with the reactor power. 
This result strongly disfavors a possible explanation in terms of neutrino oscillations as well as other new physics. 
This unexpected feature of the spectrum sheds doubt on the validity of the flux calculation as well as 
our understanding of the reactor spectra.

Radiochemical experiment using Gallium nuclei was performed by  GALLEX\cite{Kaether:2010ag} and SAGE\cite{Abdurashitov:2005tb} collaborations and
observed  a deficit of  $\nu_e$ from high intensity radioactive sources.
The weighted average value of the ratio, $R$, of measured to expected events rates of the Gallium radioactive source experiments \cite{Abdurashitov:2005tb} is 
\begin{eqnarray}
R=0.88\pm0.05,
\end{eqnarray}
which represents a deviation from unity of more than $2\sigma$, named the {\it Gallium anomaly}\cite{gallium}.
The statistical significance of the deficit  fluctuates around the $3\sigma$ level slightly depending on the assumptions made on the theoretical estimate of the cross section for the process $\nu_e+{\rm ^{71}Ga} \rightarrow {\rm ^{71}Ge}+{\rm e}^{-}$. 
The deficit can be interpreted as an indication of  $\nu_e$ disappearance due to oscillations
within the short distance $L\sim 1 m$  between the source and the detector, even though
it can be explained by
 a systematic error in the ${\rm ^{71}Ge}$ extraction efficiency and/or in the theoretical estimate of the cross-section.

Both the antineutrino reactor and gallium anomalies can be interpreted as a phenomenon of $\nu_e$ disappearance driven by sterile neutrino oscillations. 
In an effective 2-flavor scheme the results can be described by a new mass-squared difference $\Delta m^2$ and an effective mixing angle $\theta$, which can be identified with
$\Delta m^2 \equiv \Delta m^2_{41}$ and $\sin^2 2\theta \equiv 4|U_{e4}|^2(1-|U_{e 4}|^2)$
in the so-called $3+1$ framework \cite{Bilenky:1996rw}.
The simultaneous explanation of both anomalies requires values of $\Delta m^2 \sim 1.7~ \mbox{eV}^2$ and $\sin^2 2\theta \sim 0.1$ \cite{Gariazzo:2017fdh}. 
The inclusion of the recent results from NEOS\cite{Ko:2016owz} tends to shift downward the best fit value of the mixing angle to $\sin^2 2\theta  \sim 0.08$ \cite{Gariazzo:2017fdh}. 
The results of NEOS deserve some further comment. A raster scan analysis made by the collaboration excludes the hypothesis of oscillations at the $90\%$ C.L.
On the other hand, the expansion of the $\Delta \chi^2$ evidences a preference for sterile neutrino oscillations at roughly $2\sigma$  in 2 d.o.f. (standard 3-flavor case disfavored at $\Delta \chi^2=6.5)$ \cite{Gariazzo:2017fdh}. 
In addition, the region of parameters identified by NEOS lies inside the region allowed by all the current SBL data\cite{Gariazzo:2017fdh}. 
We think that these two findings are intriguing and deserve more attention.
In conclusion, a common interpretation of these results may result in the existence of a neutrino at a mass scale of 1 eV or above, which could drive the oscillation solution with $\Delta m^2\sim 1 {\rm eV}^2$ favoured by LSND and escape the results of other experiments. 

\subsection{Interpretation of Neutrino Anomalies in 3+1 Scheme}
Thanks to the observation of neutrino oscillations, it became clear that the SM  should be extended so as to
generate tiny neutrino masses.
The simplest and sufficient way to endow the active neutrinos with masses is 
to add 
right-handed neutrino fields
that are singlets under the $\text{SU}(2)_{L} \times \text{U}(1)_{Y}$ symmetry.
Since they do not have standard weak interactions, they are also called sterile neutrinos.
The introduction of the right-handed neutrinos make it possible to allow Majorana mass
$\nu_{s R}^{T} \mathcal{C}^{\dagger} \nu_{s R}$ with the charge conjugate operator $\mathcal{C}$ that are invariant under $\text{SU}(2)_{L} \times \text{U}(1)_{Y}$
gauge transformations.
When the mass scale of right-handed neutrinos is very  large compared with Dirac neutrino masses, 
it is possible to use so-called seesaw mechanism  that produces naturally tiny neutino masses for the left-handed active neutrinos.
Such heavy right-handed neutrinos are decoupled from the accessible low-energy physics
and thus do not have any impact on phenomenology.
But, not all of the right-handed neutrinos need to be very heavy, some of them could be so light that
they could be invoked in low-energy physics such as neutrino oscillations and play important roles in cosmology.

We assume that light sterile neutrino with a mass of order 1 eV interacts through mixing with the active neutrinos, the so-called $3+1$ scheme,
which is the simplest framework to tackle the neutrino anomalies observed in the SBL experiments. 
In this framework, the four massive neutrino fields $\nu_{k L}$, with $k=1,\ldots,4$,
are obtained from the active and sterile flavor neutrino fields
through an unitary transformation that diagonalizes the neutrino mass matrix \cite{Gariazzo:2015rra} :
\begin{align}
\null & \null
\nu_{\alpha L}
=
\sum_{k=1}^{4}
U_{\alpha k}
\nu_{k L}
\quad
(\alpha=e,\mu,\tau)
,
\label{mix1}
\\
\null & \null
(\nu_{s R})^{C}
=
\sum_{k=1}^{4}
U_{4 k}
\nu_{k L}
,
\label{mix2}
\end{align}
where $U$ is an unitary $4 \times 4$ matrix.
In general, the unitary $4\times 4$ mixing matrix is parameterized in terms of 6 angles and  6 phases.

In the framework of 3+1 active-sterile mixing, the SBL neutrino experiments are sensitive only to the oscillations caused by the squared-mass difference $\Delta m^2_{41}\simeq \Delta m^2_{42} \simeq \Delta m^2_{43} \sim O(1)\mbox{eV}^2$, with
$\Delta m^2_{jk}\equiv m^2_j - m^2_k$, that is much larger than the solar squared-mass difference
$\Delta m^2_{\rm SOL}=\Delta m^2_{21}$  and the atmospheric squared-mass difference $\Delta m^2_{\rm ATM} =|\Delta m^2_{31}|=|\Delta m^2_{32}|$, which generate the observed solar, atmospheric and long-baseline neutrino oscillations explained by the standard three-neutrino mixing \cite{mass-dif}. 
The unitary $4\times 4$ mixing matrix is constructed so that the unitary $3\times 3$ mixing matrix is 
extended to an unitary $4\times 4$ matrix  with $|U_{e4}|^2,|U_{\mu 4}|^2,|U_{\tau 4}|^2<< 1$. 
In the approximation, the flavor oscillation probabilities of the neutrinos in the SBL experiments are given by \cite{3+1-prob}
\begin{eqnarray}
P^{({\rm SBL})}_{\alpha \beta}\simeq \sin^2 2\theta_{\alpha \beta} 
                          \sin^2\left( \frac{\Delta m^2_{41}L}{4E}\right), ~~~
P^{({\rm SBL})}_{\alpha \alpha}\simeq 1-\sin^2 2\theta_{\alpha \alpha} 
                          \sin^2\left( \frac{\Delta m^2_{41}L}{4E}\right), \label{prob-SBL}
\end{eqnarray}
where $\alpha, \beta = e, \mu, \tau, s$, $L$ is the source-detector distance and $E$ is the neutrino energy.
The oscillation amplitudes depend only on the absolute values of the elements in the forth column of the mixing matrix,
\begin{eqnarray}
\sin^2 2 \theta_{\alpha \beta} = 4|U_{\alpha 4}|^2|U_{\beta 4}|^2, ~(\alpha \neq \beta),~~~
\sin^2 2 \theta_{\alpha \alpha} = 4|U_{\alpha 4}|^2(1-|U_{\alpha 4}|^2).
\end{eqnarray}
%
%
Since $|U_{\alpha 4}|^2$ with $\alpha=e,\mu,\tau$ is supposedly small,  the oscillation amplitudes for 
the disappearance modes are related to those for the appearance modes such as \cite{Bilenky:1996rw,tension-formular}
\begin{eqnarray}
\sin^2 2 \theta_{\alpha \beta}\simeq \frac{1}{4} \sin^2 2\theta_{\alpha \alpha}\sin^2 2\theta_{\beta \beta}.
\label{tension}
\end{eqnarray}
This relation may serve as a check if the interpretation of the disappearance modes is in consistent with
that of the appearance ones in terms of neutrino oscillations containing sterile neutrino.

\subsubsection{Combination of ~$\protect\nua{e}$ Appearance Experiments}
\begin{figure}
  \centering
  \begin{tabular}{cc}
    \includegraphics[width=.48\textwidth]{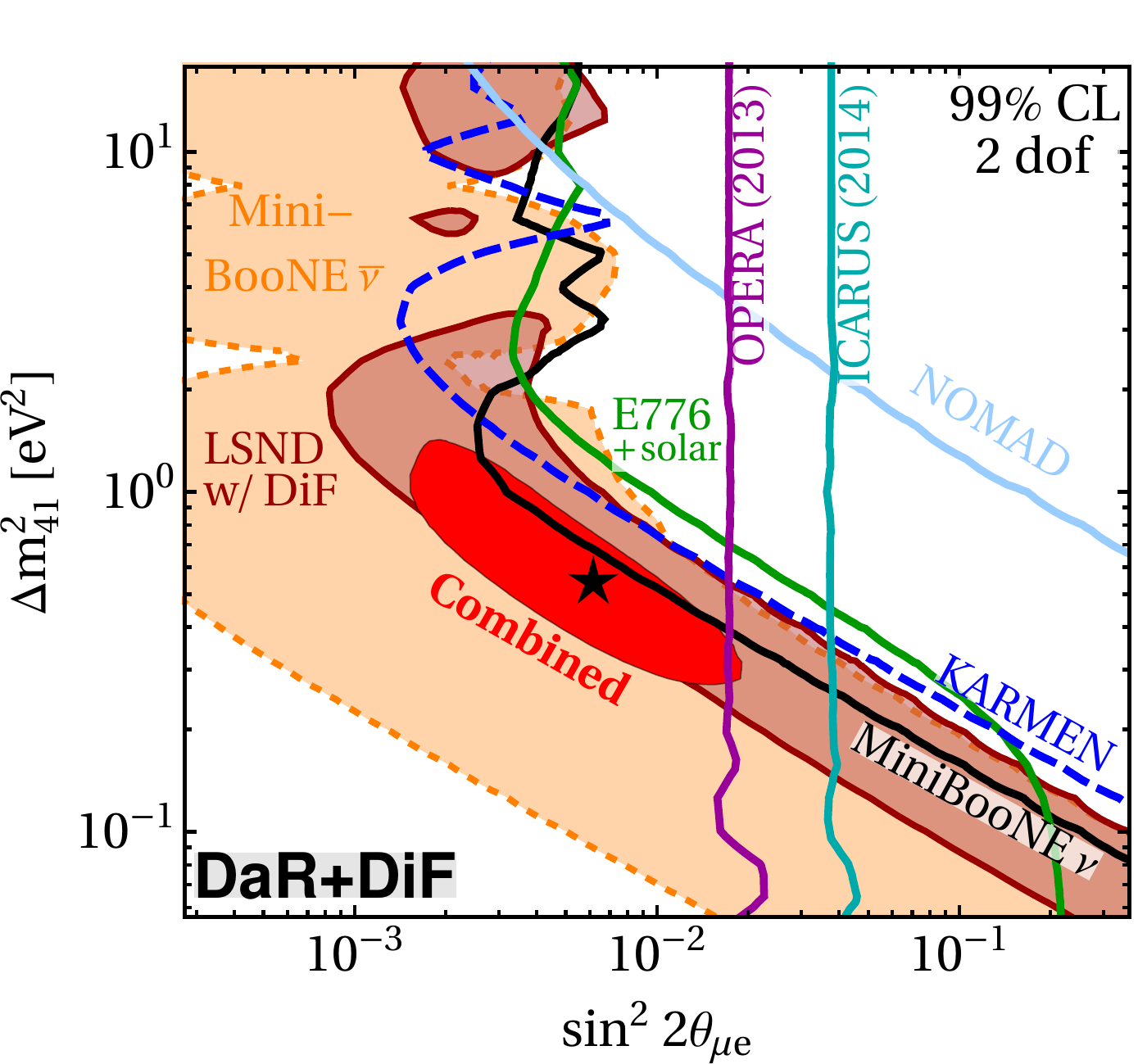}
  \end{tabular}
  \caption{Constraints on $\protect\nua{\mu} \to \protect\nua{e}$ appearances from SBL experiments in the $3+1$ scenario.
    The allowed parameter regions are projected onto the $\sin^2 2\theta_{\mu e} -
\Delta m_{41}^2$ plane \cite{Dentler:2018sju}. The red region is allowed by fit to the combined appearance data.
}
  \label{fig:nu-mu-to-nu-e}
\end{figure}
Fig. \ref{fig:nu-mu-to-nu-e} shows the combined results of $\protect\nua{e}$ appearance searches from the most global
fit projected onto the $\sin^2 2 \theta_{\mu e}-\Delta m^2_{41}$ plane\cite{Dentler:2018sju}.
In the analysis, the LSND data from both pions decaying at rest and in flight are included.
One can see that all the data sets are consistent among each other.
All the parameter space inside the colored regions is allowed at $99\%$ C.L. by two experiements
LSND and MiniBooNE that observed an excess in  $\protect\nua{e}$ appearance.
All the parameter space to the right of the colored lines is disfavored at $99\%$
C.L. by experiments that did not observed any significant excess.
No-oscillation hypothesis for all appearance data is shown to be disfavored
compared with the best fit marked by star.
But, we note that the global analysis for $\protect\nua{e}$ appearance experiments has a relatively
poor goodness of fit\cite{Dentler:2018sju} .

\subsubsection{Combination of ~$\protect\nua{e}$ Disappearance Experiments}
\begin{figure}
  \centering
   \includegraphics[width=.55\textwidth]{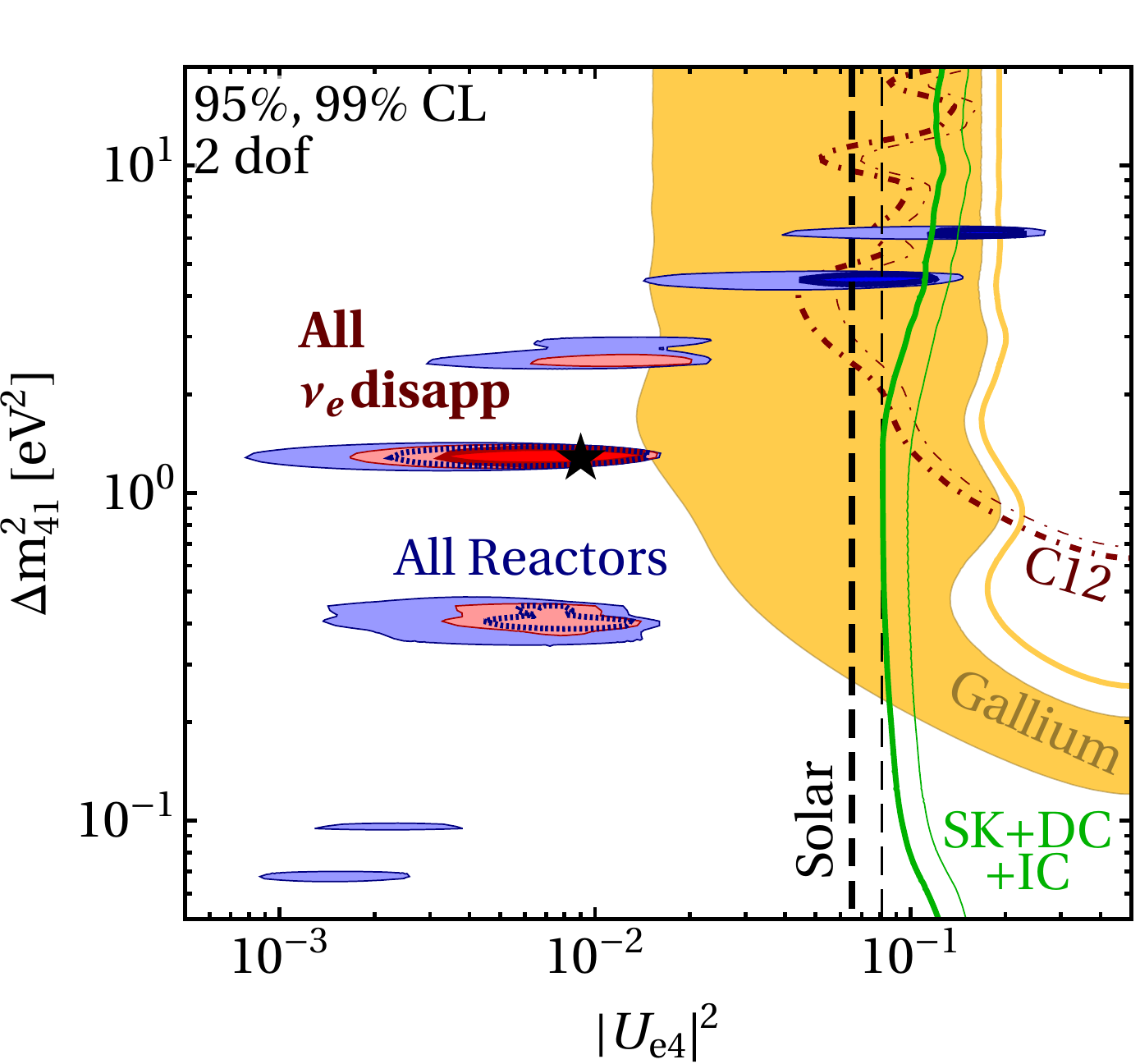}
  \caption{Constraints on $\protect\nua{e}$ disappearance in the
    $3+1$ scenario, assuming free reactor flux normalizations.  The preferred parameter regions at 95\%
    and 99\%~C.L. are projected onto the $|U_{e4}|^2$-$\Delta
    m_{41}^2$ plane\cite{Dentler:2018sju}.  The regions inside the shaded areas and to the
    right of the exclusion curves are allowed.
The green curves correspond to the limits on
    $|U_{e4}|^2$ obtained from atmospheric neutrino data from SuperK,
    IceCube and DeepCore. Figure is extracted from Ref.\cite{Dentler:2018sju}.
}
  \label{fig:nu-e}
\end{figure}
In ~$\protect\nua{e}$ disappearance channels, the most stringent constraints on the mass-mixing parameter space
come from reactor antineutrino experiments at SBL, $L\lesssim 1 km$.
Fig. \ref{fig:nu-e} shows constraints on $\protect\nua{e}$ disappearance in the $3+1$ scheme.
The preferred parameter regions at $95\%$ and $99\%$ C.L. are projected onto 
 the $|U_{e4}|^2-\Delta m^2_{41}$ plane.
The shaded regions and the parameter space to the left of the exclusion curves from solar and atmospheric data are allowed.
For the reactor analysis, free flux normalizations are assumed.
The red region includes all the disappearance data and the best fit is marked by star.
The global  $\protect\nua{e}$ disappearance data show a robust indication in favor of sterile neutrino
at $3\sigma$ C.L.

%
%
\subsubsection{Appearance - Disappearance Tension}
\begin{figure}[!ht]
\begin{center}
\begin{tabular}{cc}
\includegraphics[width=0.5\linewidth]{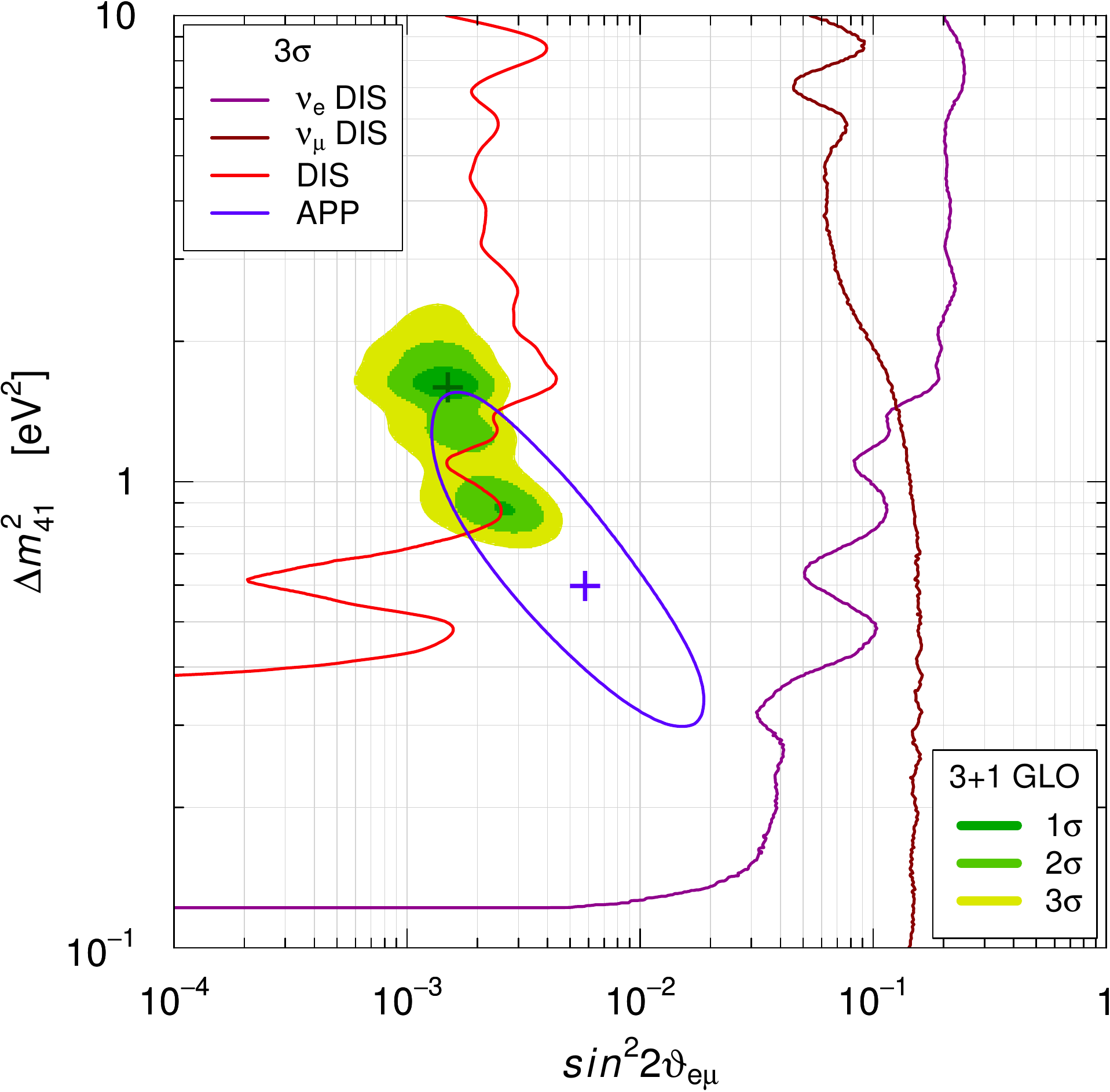}
\end{tabular}
\end{center}
\caption{Allowed regions projected onto the
$\sin^{2}2\vartheta_{e\mu}$--$\Delta{m}^{2}_{41}$
plane
obtained from the 3+1 global fit
of SBL data (GLO)
compared with the $3\sigma$ allowed regions
obtained from
appearance data (APP)
and the $3\sigma$ constraints obtained from
$\protect\nua{e}$ disappearance data ($\nu_{e}$ DIS),
$\protect\nua{\mu}$
 disappearance data ($\nu_{\mu}$ DIS)
and the
combined disappearance data (DIS).
The best-fit points of the GLO and APP fits are indicated by crosses.
Figure is extracted from Ref.\cite{Giunti:2016oan}}
\label{fig:fig1-a}
\end{figure}
We see from Eq.(\ref{tension}) that the bounds on $|U_{e4}|$ and $|U_{\mu 4}|$
from $\nu_e$ and $\nu_{\mu}$ disappearance data lead to a quadratic suppression of 
$\sin^2 2\theta_{e\mu}$ which corresponds to the amplitude of $\nu_{\mu} \rightarrow \nu_e$
oscillation.
Although  no obvious anomaly has been observed in $\nu_{\mu}$ disappearance channel,
the limits on  $|U_{\mu4}|$ can be used to check the consistency of the relation Eq.(\ref{tension}).
Fig.~\ref{fig:fig1-a} shows the allowed regions  in the
$\sin^{2}2\vartheta_{e\mu}$--$\Delta{m}^{2}_{41}$
plane
obtained from the global fit to SBL neutrino oscillation data
in the $3+1$ framework, which
compared with the bounds from the disappearance data\cite{Giunti:2016oan}.
The best-fit values of the oscillation parameters\cite{Giunti:2016oan} are given as
$\Delta{m}^{2}_{41} \sim 1.6 \, \text{eV}^2$,
$|U_{e4}|^2 \sim 0.028$,
$|U_{\mu4}|^2 \sim 0.014$,
from which those of oscillation amplitudes are determined to be 
$\sin^{2}2\vartheta_{e\mu} \sim 0.0015$,
$\sin^{2}2\vartheta_{ee} \sim 0.11$ and
$\sin^{2}2\vartheta_{\mu\mu} \sim 0.054$.
%
The purple  ($\nu_e$ DIS) and maroon ($\nu_\mu$ DIS) curves represent  the $3\sigma$ exclusion obtained
from $\protect\nua{e}$ and
$\protect\nua{\mu}$  SBL disappearance data, respectively, by
imposing the relation, $\sin^{2}2\vartheta_{e\mu}
\leq
4 |U_{e(\mu)4}|^2 \left( 1 - |U_{e(\mu)4}|^2 \right)
=
\sin^{2}2\vartheta_{ee(\mu \mu)}$.
The red curve (DIS) corresponds to  the $3\sigma$ exclusion from the combined
SBL disappearance data by applying the relation Eq.(\ref{tension}).
The region surrounded by the blue curve (APP) is
the $3\sigma$ allowed region obtained from $\protect\nua{\mu}\to\protect\nua{e}$
SBL appearance data.
While the separate $3\sigma$ exclusion curves for $\protect\nua{e}$ and
$\protect\nua{\mu}$  SBL disappearance modes do not exclude
the  $3\sigma$  allowed region for the
$\protect\nua{\mu}\to\protect\nua{e}$ SBL appearance model,
most of the region surrounded by blue curve is excluded by the one obtained  the combined
SBL disappearance data .
Thus, this indicates a strong
appearance-disappearance tension
\cite{Gariazzo:2015rra,Giunti:2011gz}.

Recent analysis performed in Ref.\cite{Dentler:2018sju} concludes that the sterile neutrino oscillation hypothesis
as an explanation of the LSND and MiniBooNE anomalies  is strongly disfavored, whereas it remains
a viable option for the reactor antineutrino and Gallium anomalies.
Such a conclusion comes out from the strong tension between the anomalies in the appearance data
and disappearance data, which is supported by rather severe constraints on the mixing parameter
in $\protect\nua{\mu}$ disappearance mode due to recent data from MINOS/MINOS+ \cite{Adamson:2017uda} and IceCube
\cite{TheIceCube:2016oqi}.  

An interesting proposal to alleviate the tension has been suggested in  Ref.~\cite{Giunti:2013aea}.
The key idea of the proposal is to exclude from the fit  an anomalous excess of $\nu_{e}$-like events 
appeared in the low-energy bins of the MiniBooNE experiment
\cite{AguilarArevalo:2008rc,AguilarArevalo:2013pmq}.
Since the MiniBooNE low-energy excess requires a small value of $\Delta{m}^2_{41}$
and a large value of $\sin^22\vartheta_{e\mu}$
\cite{Giunti:2011hn}, 
exclusion of the low-energy excess from the fit makes the allowed region by appearance data
shifted towards larger values of $\Delta{m}^2_{41}$
and smaller values of $\sin^22\vartheta_{e\mu}$.
As a result, the overlap of  the region allowed by both appearance and disappearance data gets increased.
%

\subsection{Impact of light sterile neutrino on neutrinoless double beta decay}
%
The existence of sterile neutrino with a mass of eV scale has important implications on the
neutrinoless double beta  (\nbd) decay through the effective neutrino mass \cite{Bilenky:1999wz,Goswami:2005ng,Goswami:2007kv,Barry:2011wb,Li:2011ss,Girardi:2013zra,Giunti:2015kza,Liu:2017ago,Jang:2018zug}.
As is well known \cite{Schechter:1981bd,Duerr:2011zd,Liu:2016oph,Rodejohann:2011mu}, the \nbd~ decay violates the lepton number by two units, and naturally happens when $\protect\nua{e}$  is an admixture of mass eigenstates of  Majorana neutrinos.
Allowing the sterile neutrino with a mass of eV scale mixing with the active neutrinos, the \nbd~ decay rate is
proportional to the effective neutrino mass given as \cite{Goswami:2005ng}
\begin{equation}
\langle m \rangle_{ee}\equiv \left|\sum^4_{i=1}m_iU^2_{ei}\right|=|
m_1 c^2_{12}c^2_{13}c^2_{14}+m_2 e^{i\alpha} c^2_{13} c^2_{14} s^2_{12}+
m_3 e^{i\beta} c^2_{14}s^2_{13}+m_4 e^{i\gamma} s^2_{14}|, \label{effective-mass}
\end{equation}
where $m_i$ and $U_{ei}$ stand for the individual neutrino masses and the first-row entries of the
$4 \times 4$ neutrino mixing matrix, respectively, and $c_{ij}=\cos \theta_{ij}, s_{ij}=\sin\theta_{ij}$.
We see that the effective neutrino mass is sensitive to the Majorana phases, $\alpha, \beta$, and $\gamma$, whereas the Dirac phases do not appear in $\langle m \rangle_{ee}$.

It is worthwhile to note that in the framework of three active neutrinos the effective neutrino mass can vanish for the normal mass hierarchy, whereas it can not vanish for the inverted mass hierarchy  \cite{Pascoli:2002xq,
Benato:2015via,Xing:2016ymd,Ge:2016tfx}
However, the effective neutrino mass given by Eq.(\ref{effective-mass}) can be zero irrespective of the
mass hierachy of three active neutrinos \cite{Liu:2017ago}.
This feature implies that neutrinos can be Majorana particles even in the case that the
\nbd~decay  is absent for the inverted mass hierarchy of three active neutrinos.
In Ref. \cite{Liu:2017ago}, it has also been shown by taking $\Delta m^2_{41}=1.7 {\rm eV}^2$ and
$\sin^2\theta_{14}=0.019$ corresponding to the best fit extracted from Ref.\cite{Gariazzo:2017fdh}
 that the contributions from the sterile neutrino in the $3+1$ scheme are comparable to those from three active neutrinos in both normal and inverted mass hierarchical cases,  implying an intriguing interplay between them.
%
%

\section{Light Sterile Neutrino in Cosmology}
Sterile neutrinos can play an important role in the evolution of the  Universe, modifying the cosmological observables.
In this section we review on the aspects of sterile neutrinos in cosmology and we discuss how
the analysis of current cosmological observations can be used to learn about sterile neutrino properties such as a mass and mixing
with active neutrinos with very good perspectives from future cosmological measurements. 
To see how cosmology is sensitive to the possible existence of sterile neutrinos, we first consider
detectable imprints of sterile neutrinos on cosmological observables that can be used to constrain
properties of sterile neutrinos.
In previous section,  we have considered a sterile neutrino with a mass of eV scale motivated by
a  few anomalies in the SBL neutrino oscillation data. 
Those neutrinos can be produced by oscillations of the active neutrinos in the early universe, leaving possible traces on different cosmological observables.
The contribution of the light sterile neutrino depends on whether it is (was) relativisitic or not in some epoch in early Universe.
If those sterile neutrinos are produced well before the active neutrino decoupling, they aquire quasi-thermal distributions and behave as extra degrees of freedom at the time of BBN.
Let us begin by studying the relavent observables that can be useful to probe the existence of such a light sterile neutrino with a mass of eV scale.
And then, we move to review on the cosmological roles of sterile neutrino with a mass of keV scale. For such a neutrino, we will mainly discuss the possibility
that it can be DM candidate and play a key role in achieving low scale leptogenesis.

\subsection{Parameters and Constraints}
\paragraph{Parameters} : 
To probe imprint of light sterile neutrino on cosmological observables, let us first introduce two parameters whose constraints or values can be extracted from cosmological data.
The first one is the so-called {\it effective number of relativistic species} and the other is {\it effective sterile neutrino mass}.
The effective number of relativistic species, $\Neff$, is  responsible for the total energy density in radiation
excluding photons
and  defined by \cite{Shvartsman:1969mm,Steigman:1977kc}
\begin{equation}
\Neff \equiv \frac{\rho_{r}-\rho_{\gamma}}{\rho^{\thm}_{\nu}}, \label{Neff}
\end{equation}
where $\rho^{\thm}_{\nu}=(7 \pi^2/120)(4/11)^{4/3}T^4_{\gamma}$ is
the energy density of one SM massless neutrino with a thermal distribution,
$\rho_{\gamma}$ is the energy density of photons, and $\rho_r$ is the total energy density in relativistic particles.
The expression (\ref{Neff}) is valid when relativistic particles contributing to $\Neff$ are decoupled from the thermal plasma  at $T\sim 1$ MeV below which photons, electrons and positrons are in thermal equilibrium.
In the SM, by the time of BBN only the three known neutrino species contribute to $\rho_r$,
resulting in $\Neff=3.046$\cite{Mangano:2005cc}.
This is slightly larger than three because the SM neutrinos are not fully decoupled at $e^+e^-$ annihilation, they do share some of the energy (entropy) when the  $e^+e^-$ pairs annihilate \cite{Mangano:2005cc}.
Extra radiation beyond the SM would cause a deviation from above the SM value of $\Neff$, which
is denoted as $\DNeff$.
Although adding an extra light fermion could contribute $\DNeff=1$, most generally 
$\Neff$ is noninteger and varies with time.
The presence of a non-zero $\DNeff$ due to sterile neutrinos would modify the cosmic expansion rate and affect the BBN of light elements, the CMB anisotropies, and the formation of LSS.
We will present the constraints on $\DNeff$ obtained from those observations later.

Massive neutrinos become non-relativistic when their masses are equal to their average momentum, given for any Fermi-Dirac-distributed particle by $<p> = 3.15T$. 
Thus the redshift of the non-relativistic transition is given by \cite{Lesgourgues:2012uu}
$z^{\rm nr}_i = m_{\nu_i}/(3.15T^0_{\nu} )-1$  
for each eigenstate of mass $m_{\nu_i}$.
The sterile neutrinos with a mass between $10^{-3}$ eV and 1 eV can contribute not only to the radiation density at the time of equality and but also to  the non-relativistic matter density today.
Assuming that the light sterile neutrino has a thermal distribution with an arbitrary temperature
$T_s$,  the true mass of a thermally distributed sterile neutrino reads
\begin{equation}
m^{\thm}_s\equiv \left(\frac{T_{\nu}}{T_s}\right)^3 \meff{s}=\DNeff^{-3/4} \meff{s},
\end{equation}
where the sterile neutrino distribution is assumed to be the same as for the active neutrinos in
the instantaneous decoupling limit.
The sterile neutrino contribution to the  non-relativistic matter energy density can be parameterized
in terms of the dimensionless parameter~\cite{Acero:2008rh}
\begin{equation}\label{eq:omegas}
\omega_{s}=\Omega_{s} h^2=\frac{\rho_{s}}{\rho_{\rm c}} \, h^2
=\frac{h^2}{\rho_{\rm c}}\frac{m_{s}}{\pi^2}\int dp \, p^2 f_{s}(p),
\end{equation}
where
$\rho_{s}$ and $\rho_{\rm c}$ are the energy density of non-relativistic sterile neutrinos and
critical density, respectively. 
Alternatively, $\omega_{s}$ can be converted to the effective sterile  neutrino mass, $\meff{s}$ through the relation \cite{Ade:2013zuv,Ade:2015xua} 
\begin{equation}\label{eq:meffs}
\meff{s} = 94.1 \, \omega_{s} \, \mathrm{eV}.
\end{equation}
In this case, the total neutrino density becomes $\Omega_{\nu} h^2 = 0.00064+\Omega_{\nu_s}h^2$, where 0.00064 is the contribution to the neutrino density by the three active neutrinos, $\Omega_{\nu} = \frac{\sum m_{\nu_i}}{93.03 h^2{\rm eV}} $, assuming the sum of three neutrino masses to be $\sum m_{\nu_i} = 0.06$ . Thus, any change from $\Omega_{\nu} h^2 = 0.00064$ may be a hint of the existence of light sterile neutrino. 

Massive neutrinos could leave distinct signatures on the CMB and LSS at different epochs of the Universe's evolution \cite{Abazajian:2014gza}. To a large extent, these signatures could be extracted from the available cosmological observations, from which the total neutrino mass could be constrained. Currently, the CMB power spectrum, combined with LSS and cosmic distance measurements can provide tight limits on the total mass of neutrinos \cite{Ade:2013zuv,Ade:2015xua}. 
If the production is non-thermal, instead, there are several possible scenarios. A popular one is the non-resonant production or Dodelson-Widrow (DW) scenario \cite{Dodelson:1993je}. In the DW scenario 
the momentum distribution function of light sterile neutrino can be written as $f^{DW}_s (p) = \beta (e^{p/T_{\nu}}+ 1)^{-1}$, where $\beta$ is a normalization factor, and one obtains $\meff{s} = \DNeff m_s$.
%
%
%
%
%
%
\paragraph{Constraints for Relativisitic Sterile Neutrino} : 
The relativistic sterile neutrino can affect the expansion rate during the BBN, which
in turn affects the observed abundance of light elements and in particular that of $^4{\rm He}$.
From the observed primordial abundance of helium and other light nuclear species, one can obtain an estimate of the energy density of the Universe at the epoch of BBN, that can be translated in an estimate of $\Neff$ 
, which can then lead us to extract an upper limit: \cite{Steigman:2012ve,Iocco:2008va,Jacques:2013xr,Fields:2014uja}
\begin{equation}\label{BBN-Neff}
 \Neff<3.3 .
\end{equation} 
In the absence of mixing, sterile neutrinos that interact only extremely weakly will not be in thermal equilibrium 
at the nucleosynthesis epoch, and therefore the limit (\ref{BBN-Neff}) cannot in general exclude their existence.
However, mixing between active and sterile neutrinos induces oscillations that can generate a significant population of sterile neutrinos. If the mixing is sufficiently large and the oscillations sufficiently fast,  one can violate the bound (\ref{BBN-Neff}). Therefore the bound (\ref{BBN-Neff}) can be used to set limits on the existence of oscillations between active and sterile neutrinos. 
According to Ref.~\cite{Mangano:2011ar}, BBN constrains $\Neff$ such that
$\DNeff<1$ at 95\% C.L., regardless of the inclusion of CMB constraints on the baryon density $\Omega_b h^2$.
In addition, a non-zero value of $\DNeff$  due to sterile neutrinos
could affect the evolution of the  CMB anisotropies in a few ways, as studied in\cite{ Lesgourgues:2014zoa}.

We see that the Planck satellite measurement is in very good agreement with both the standard prediction of $\Neff = 3.046$ and BBN results, in spite of a marginal preference for extra relativistic degrees of freedom (exacerbated if astrophysical measurements of $H_0$ are included)\cite{Ade:2015xua}.
More recently, the authors of Ref.~\cite{Cyburt:2015mya} obtained
$\DNeff<0.2$ at 95\% C.L.
by taking into account the BBN and CMB data.
In the analysis of the Planck collaboration \cite{Ade:2015xua}
the constraints on $\Neff$ alone have been obtained considering massless neutrinos in the \lcdm+$\Neff$ model.
However, it is interesting to present them in order to understand how the constraints on $\Neff$ change
when different experimental results are taken into account.
Considering the temperature (Planck TT) and the low-$\ell$ polarization (lowP) data can result in
$\Neff=3.13\pm0.32$
\cite{Ade:2015xua},
which is consistent with the standard three-neutrino value $\Neff=3.046$.
The inclusion of BAO observations \cite{Anderson:2013zyy, Ross:2014qpa, Beutler:2011hx} tightens slightly the constraint to $\Neff=3.15\pm0.23$,
leading to the upper bound $\DNeff<1$ at more than $3\sigma$
\cite{Ade:2015xua}.
Probing $\DNeff$ precisely at different epochs, during BBN and at the formation of the CMB,
can make it possible to discriminate among different models.

\begin{figure}[t]
\centering
\includegraphics[width=92mm]{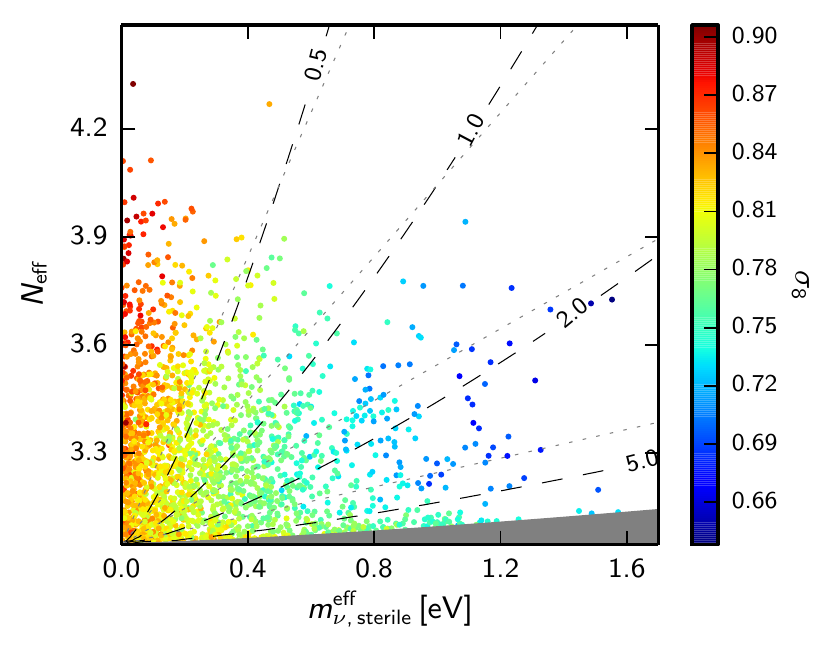}
\caption{Samples from Planck temperature data in the
$\Neff$--$m_{\rm sterile}^{\rm eff}(\equiv m^{\rm eff}_s)$ plane, 
colour-coded by $\sigma_8$ (late-time fluctuation amplitude), for models
with one massive sterile neutrino , 
and three active neutrinos as in the $\Lambda$CDM model \cite{Ade:2015xua}.
The physical mass of thermally produced sterile neutrino,
$m_{s}^{\rm th}$, is constant along the grey dashed lines, with
the indicated mass in $\mathrm{eV}$; the grey shaded region is
excluded by  $m_{s}^{\rm th}< 10$ eV,
which cuts out most of the area where the neutrinos behave nearly like
DM. The dotted lines correspond to the fixed physical masses in the Dodelson-Widrow
scenario, $m_{s}^{\rm DW}$.}
\label{fig:meffsterile-sigma8}
\end{figure}

%
%
%

\paragraph{Constraints for Non-relativistic Sterile Neutrino} : 
Here, we consider the light sterile neutrinos which are counted as radiation at the time of radiation-matter equality and as non-relativistic matter today.
As mentioned above, the sterile neutrinos with a mass in the range from $10^{-3}$ eV to $1$ eV.
The non-relativistic massive sterile neutrinos may affect the CMB temperature and polarization spectrum shape  in a few ways as discussed in \cite{Lesgourgues:2012uu}.
Since both the redshift of matter-to-radiation equality $z_{\rm eq}$ and baryon density
are well-constrained parameters, matter density $\omega_m$ and $\Neff$ through $z_{\rm eq}$ is exactly degenerate, due to which the CMB alone is not a very powerful tool for constraining sub-eV
neutrino masses.
In order to break the $\Neff$-$\omega_m$ degeneracy, it is desirable to combine the CMB  data
with  homogeneous cosmological constraints and/or
measurements of the LSS power spectrum and so on \cite{Lesgourgues:2012uu}.
The LSS of the Universe can be probed by the measurement of the matter power spectrum at a
given redshift $z_{\rm eq}$.
Since the shape of the matter power spectrum is affected by the free-streaming caused by 
massive neutrinos with a mass of the eV scale, it is the key observable to constrain
neutrino masses.

The Planck collaboration has released results for sterile neutrino by assuming that
two active neutrinos are massless, one active neutrino has a mass of 0.06 eV
and one sterile neutrino has a mass $m_s$\cite{Ade:2015xua}.
In the analysis, they take a phase-space distribution of the sterile neutrino equal to the one of active neutrinos multiplied by a suppression factor $\chi_s$, which corresponds to the generation of sterile neutrinos
through non-resonant oscillations proposed in \cite{Dodelson:1993je}.
The authors reported the result of a search for sterile neutrino with the latest cosmologcial observations\cite{Feng:2017nss}.
In their analysis, what they include are the Planck 2015 temperature and polarization data\cite{Ade:2015xua}, the baryon acoustic oscillation (BAO) data \cite{Cuesta:2015mqa,Ross:2014qpa, Beutler:2011hx}, the Hubble constant ($H_0$) direct measurement \cite{Riess:2016jrr}, the Planck Sunyaev-Zeldovich (SZ) cluster counts data\cite{Ade:2015fva}, the Planck lensing data and the cosmic shear data of weak lensing (WL) \cite{Ade:2015zua} from the CFHTLenS survey \cite{Heymans:2013fya}.
They consider $\Lambda$CDM+$\Neff$+$m^{\rm eff}_s$ model by using two data combinations,
one is CMB+BAO and the other is CMB+BAO+other.
The constraints on $\Neff$ and $m^{\rm eff}_s$ are obtained  at $95.4\%$ C.L. as follows\cite{Feng:2017nss} :
$$
\left.
\begin{array}{c}
N_{\rm eff}< 3.4273 \\
m_{s}^{\rm eff}< 0.7279~ {\rm eV}~
\end{array}
\right\} \quad\mbox{CMB+BAO},
$$
$$
\left.
\begin{array}{c}
N_{\rm eff}= 3.30^{+0.12}_{-0.20} \\
m_{s}^{\rm eff}< 0.2417~ {\rm eV}~
\end{array}
\right\} \quad\mbox{CMB+BAO+other}.
$$
The results shows that $N_{\rm eff}$ cannot be well constrained using only the CMB+BAO data, but the addition of $H_0$, SZ, Lensing, and WL data can significantly improve the constraint on $N_{\rm eff}$, favoring $\Delta N_{\rm eff}>0$ at the 1.27$\sigma$ statistical significance.  Evidently, adding low-redshift data tightens the constraint on $m_s^{\rm eff}$ significantly\cite{Feng:2017nss}. This indicates that the SZ cluster data (as well as the $H_0$, Lensing, and WL data) play an important role in constraining the mass of sterile neutrino.
This result seems to favor a massless sterile neutrino, in tension with the previous SBL neutrino oscillation experiments that prefer the mass of sterile neutrino at around 1~eV.

As shown in Fig. \ref{fig:meffsterile-sigma8}, in spite of the stringent constraint on $m^{\rm eff}_s$
given above, a sterile neutrino with a mass of 1 eV is possible if $\DNeff$ is quite small.
However,  sterile neutrinos introduced to resolve the SBL neutrino anomalies would imply full thermalization
before CMB decoupling in the early Universe, resulting in $\DNeff \simeq 1$.
Therefore, such a high value of $\Neff$, especially combined with $m_s \simeq 1$ eV, is strongly disfavored.

%
\subsection{Sterile Neutrino As a Dark Matter}
%
Since neutrinos in the SM are the only electrically neutral and long-lived particles, they could play a role of DM.
In order to constitute the whole DM, the sum of three neutrino masses in SM should be about 11.5 eV \cite{Vagnozzi:2017ovm},
which is clearly in conflict with the existing experimental bounds on neutrino mass as well as those from cosmological data.
Moreover, since neutrino mass is much smaller than their decoupling termperature, they become non-relativistic in the matter-dominated epoch, which 
implies that they are hot DM disfavored by current cosmological data.
Hence,
we need to extend the SM in order to accommodate DM candidate that provides the right relic abundance.
Sterile neutrinos do not carry any SM gauge charges and thus are free from any of the known forces of nature except gravity.  In order to be viable DM candidates, they must, however, be much heavier than neutrinos in the SM and have some interactions with other particles which generate a number of the population.
The sterile neutrinos with masses of a few keV can be a viable DM candidate, and are expected to leave imprints
in various cosmological observations.  We will discuss how they can produce in the early Universe and can be plausible as a DM candidate
by confronting them with the current cosmological data.
%
%
In particular, a sterile neutrino with a mass of keV scale can be a warm DM \cite{Dolgov:2000ew} that has been proposed to resolve small scale structure problems faced by cold dark matter .
How warm the sterile neutrinos are depends on how they are produced in the early Universe.
Now let us review on some mechanisms to produce sterile neutrinos with a mass of keV scale in the early Universe and then examine how they can be constrained by several cosmological data.

%
\subsubsection{Non-resonant Production via Mixing (freeze in)}
%
In models where the sterile neutrinos mix with active neutrinos such as type-I seesaw model,  the sterile neutrinos can be produced through
coherent oscillations between active and sterile states with a rate reduced by the active-sterile mixing angles
\cite{Langacker:1989sv,Dodelson:1993je,Abazajian:2001nj,Merle:2015vzu}.
 As a neutrino propagates through the plasma of the early Universe it can coherently forward scatter on particles that carry weak charge, which makes the neutrino have an in-medium mixing angle $\theta_m$
which is generally different from the vacuum mixing angle. 
Depending on the mixing angle $\theta_m$ and the oscillation phase,  a coherently propagating neutrino  has an amplitude to be either active or sterile at any point along its path.
Alternatively, the sterile neutrino can be produced through decoherent scattering processes with
overall scattering rate $\Gamma_{\nu_{\alpha}}\sim G^2_F T^5$ where $G_F$ is the Fermi constant.
After a neutrino suffers a scattering process, it propagates coherently and can be measured in a sterile state
with the propability proportional to $\sin^2 2\theta_m$, due to which the sterile neutrino production rate becomes $\Gamma_{\nu_s}\sim G^2_F T^5 \sin^2 2\theta_m$. (For the details of $\theta_m$, see \cite{Adhikari:2016bei, Boyarsky:2018tvu}.)
At low temperatures the production of the sterile neutrino is suppressed because scattering rate becomes too  small, whereas at high temperatures $\theta_m$ is suppressed due to the temperature dependent potential
that is proportional to $T^4$ and appears in the denomenator of the formular for $\sin^2 2 \theta_m$.
Due to this suppression, sterile neutrinos with mixing angle smaller than $\sin^2 2\theta \sim 10^{-6}$
were never in thermal equilibrium in the early Universe and thus allows their abundance to be smaller than the equilibrium one. Therefore, the sterile neutrino is produced via incomplete freeze in, and the particles’ momentum distribution is non-thermal, which is important in the context of cosmic structure formation.
Moreover, a sterile neutrino with these parameters is important for the physics of supernovas\cite{Fryer:2005sz}, and was proposed as an explanation of the pulsar kick velocities\cite{Kusenko:1997sp, Fuller:2003gy,Barkovich:2004jp}.
The peak rate of the sterile neutrino production through this mechanism occurs at temperature
\cite{Dodelson:1993je,Barbieri:1989ti}
\begin{equation}
T_{\rm max}\simeq 133 {\rm MeV} \left( \frac{m_s}{1 {\rm keV}}\right)^{1/3},
\end{equation}
where $m_s$ is the sterile neutrino mass.
For keV sterile neutrino DM, they are produced above temperature $T\geq 100$ MeV.

It is known that for any non-zero $\theta$, sterile neutrinos non-resonantly produced 
contributes to the relic density.
If this constribution consists of the whole observed relic density, sterile neutrinos produced by this mechanism
are excluded by a combination of structure formation bounds, X-ray searches and Lyman-$\alpha$ forest as shown in
\cite{Merle:2015vzu}.

\subsubsection{Resonant Production via Mixing}
The non-resonant production of sterile neutrinos is operative when the lepton asymmetry in the primordial plasma is negligibly small.
However, the constraints on the lepton asymmetry are significantly weaker than those on the baryon asymmetry, which opens up new possibility of sterile neutrino production.
As the Mikheyev-Smirnov-Wolfenstein (MSW) effect\cite{MSW}  enhances the transition between two active neutrinos,
the presence of a large enough lepton asymmetry in the early Universe may resonantly enhance the transition between active and sterile neutrinos.~\cite{Abazajian:2001nj,Enqvist:1990ek,Shi:1998km,Ghiglieri:2015jua}
Such a resonance could produce sterile neutrinos having a very sharp momentum at a given resonance temperature $T_{\rm res}$, which results in a characteristic peak in the distribution function on top of a spread-out contiuum part.
Since this peak is located at a comparatively small momentum, sterile neutrinos with a mass of keV scale produced resonantly are generally colder than those with the same mass produced non-resonantly shown above.
It is known that the relic density of sterile neutrinos and their energy spectrum nontrivially depend on the active neutrino scattering rates and in-medium active-sterile mixing angle\cite{Kishimoto:2008ic} . 

Numerically, accounting for all relic density with a sterile neutrino with 7.1 keV mass produced via this mechanism demands an initial net lepton asymmetry $(n_{\nu_{\alpha}}-n_{\bar{\nu}_{\alpha}})/n_{\gamma} \sim 10^{-5}-10^{-4}$~~\cite{Abazajian:2001nj, Laine:2008pg,Abazajian:2014gza} for $\sin^2 \theta \sim 10^{-11}-10^{-10}$ as suggested by the X-ray observations\cite{Bulbul:2014sua,Boyarsky:2014jta}.
Ref.\cite{Schneider:2016uqi} shows that sterile neutrino DM produced resonantly  is disfavored by the combined 
constraints from Lyman-$\alpha$ forest, dwarf galaxy number counts, and X-ray data.
Particularly, depending on how aggressively the Lyman-$\alpha$ data is interpreted, the resonant production of sterile neutrino may even be excluded\cite{Merle:2017jfn}.

\subsubsection{Production by Particle Decay}
An alternative way to generate sterile neutrino DM with a mass of keV scale is via the decay of a heavy parent particle  into sterile neutrinos \cite{Shaposhnikov:2006xi,Bezrukov:2009yw,Kusenko:2006rh,Petraki:2007gq,Merle:2013wta,Boyarsky:2018tvu} 
The parent particle can be in-equilibrium and/or our-of-equilibirum when the sterile neutrinos are produced.
There are many models to implement this mechanism to produce sterile neutrinos \cite{Adhikari:2016bei,Boyanovsky:2008nc,Roland:2014vba,Matsui:2015maa,Roland:2016gli,Frigerio:2014ifa,Kang:2014mea,Drewes:2015eoa,Adulpravitchai:2015mna,Shakya:2016oxf,Shuve:2014doa,Abada:2014zra} .
If the decay occurs after the heavy particle has frozen out or before it reaches thermal equilibrium,
which means out of thermal equilibrium decay, 
this mechanism corresponds to the non-thermal production.
The sterile neutrino produced from the decay of a heavy scalar that is a frozen-out relic  becomes cooler than that produced via mixing with active neutrinos, which is less constrained from cosmological bounds.
Regardless of the mechanism, if the active-sterile mixing is large enough for non-resonant or resonant
production via mixing to be significant, that should be included in the estimation of relic density.

The production of sterile neutrinos from the decay of bosonic particle in equilibrium 
was first calculated in a model where the inflaton plays the role of the parent particle \cite{Shaposhnikov:2006xi}.
The decay of a gauge-singlet Higgs boson at temperature of the order of the scalar mass, $T\sim 100$ GeV, can be responsible for the production of sterile neutrinos.
In this case, the  abundance of sterile neutrinos $Y_s=n_s/s$ is given in the limit that the number of degrees of freedom during the production epoch, $g_{\ast}$, is constant by\cite{Shaposhnikov:2006xi}
\begin{eqnarray}
Y_s= \frac{27\zeta(5)\Gamma }{16 \pi^3 g_{\ast}}\frac{M_0}{m_s},
\end{eqnarray}
where $\Gamma$ is the partial width for scalar decay into sterile neutrinos and
$M_0=\left(\frac{45M^2_{\rm pl}}{4\pi^3 g_{\ast}}\right)^{1/2}$ is the reduced Planck mass.
The average momentum of sterile neutrinos produced in this mechansim is
$\langle p \rangle/T=\pi^6/(378\zeta(5))T\simeq 2.45$ at $T\sim 100$ GeV, which is lower than thermal $\langle p \rangle/T\simeq 3.15$ as well as
that from non-resonant  ($\langle p \rangle/T\sim 2.83$) . 

The singlet scalar would also be so weakly coupled to the SM Higgs boson so that they could  go and decay out of equilibrium at temperature below the scalar mass.
We will consider in detail this possibility later by introducing a model including sterile neutrino with a mass of keV scale.
It is worthwhile to note that  the decay production mechanism is in the better agreement with cosmological data and  can lead to very involved spectra with two or partially even three different characteristic momentum scales, which can help to address the small scale problems of cosmic structure formation
~\cite{Merle:2013wta,dec-FIMP}.
%
\begin{figure}
\includegraphics[width=0.9\textwidth]{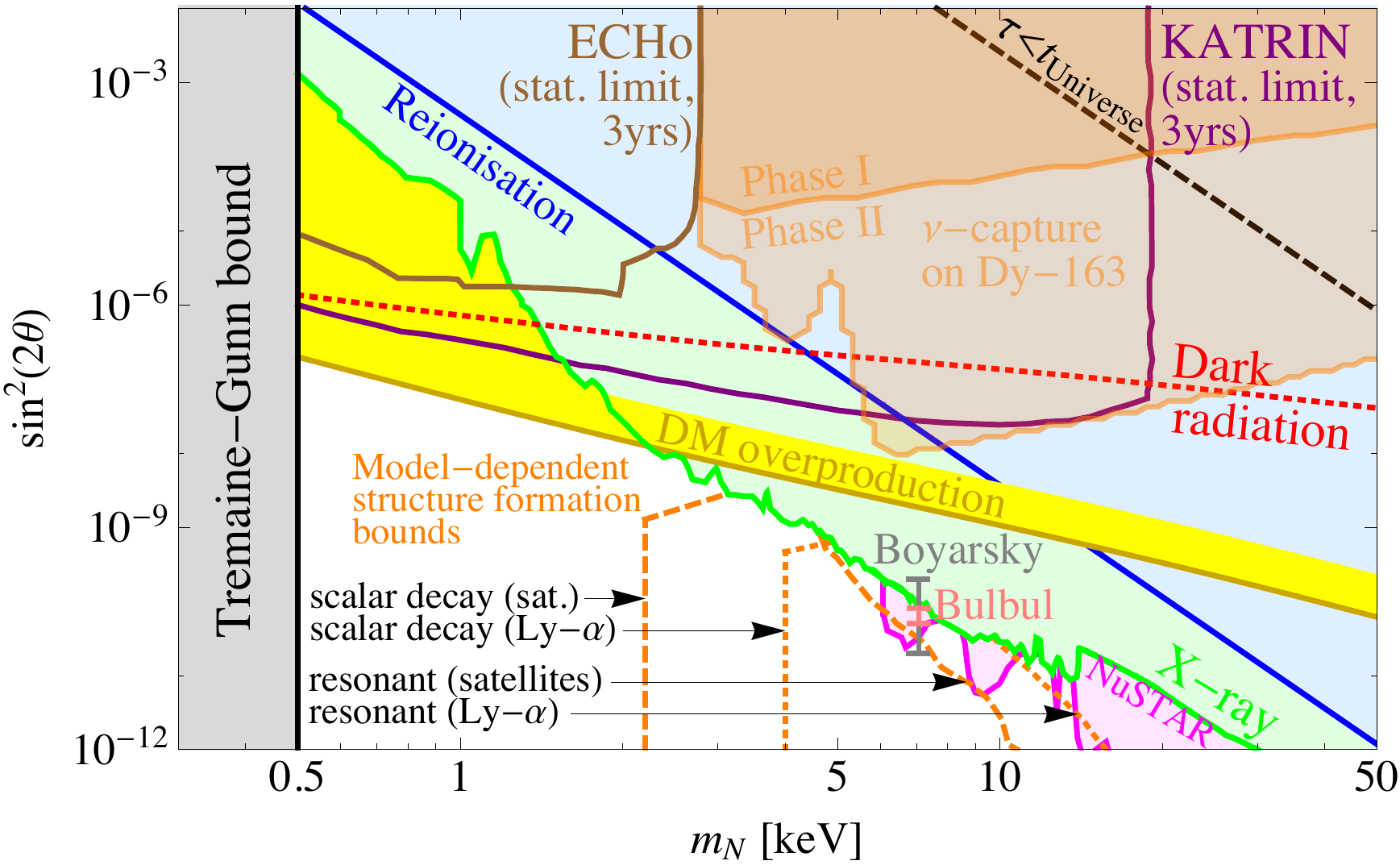}
\caption{\label{fig:Summary}Summary plot of current constraints and sensitivities of  future experimental reaches \cite{Merle:2017jfn}.}
\end{figure}

\subsubsection{Constraints on keV Sterile Neutrino}
\begin{figure}[h!]
\begin{center}
\includegraphics[width=5.0in]{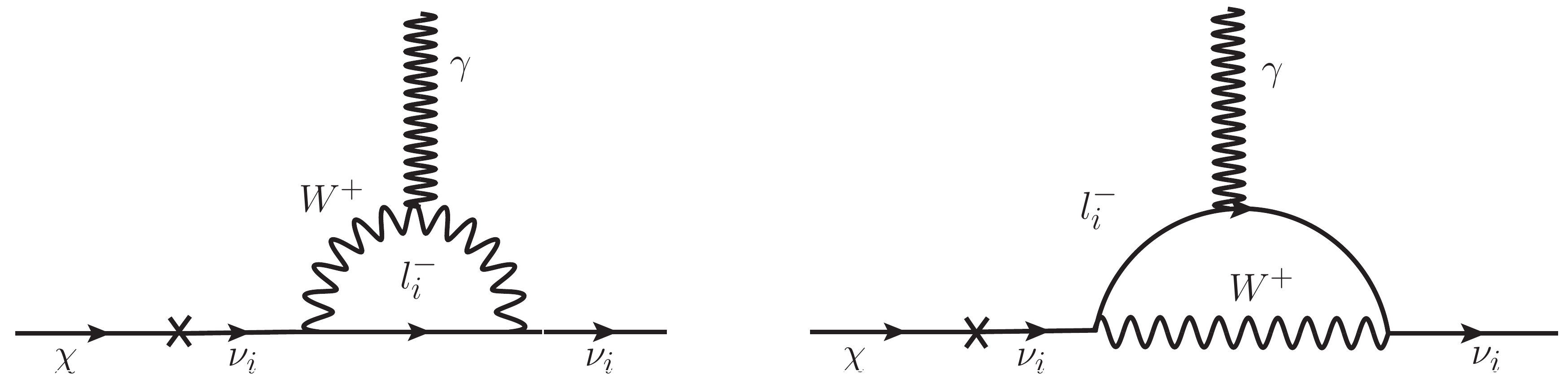}
\includegraphics[width=5.0in]{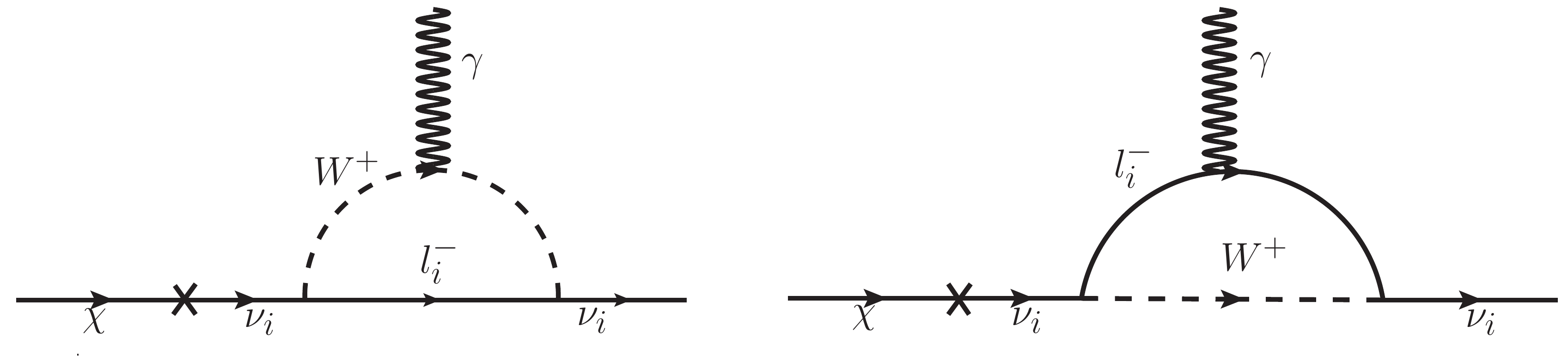}
\caption{Radiative decay of sterile neutrino DM \cite{Kang:2014mea}.}
\label{fig:nugamma}
\end{center}
\end{figure}

\paragraph{Bounds From X-ray Observations} :
 The DM particle must be stable on cosmological time scales. 
However, DM particles can decay if they have decay lifetimes longer than
the age of the Universe,
 $t_{\tiny {\rm Universe}}= 4.4\times 10^{17} s~$ \cite{Ade:2015xua},
 in all considered channels, and the decay signals may be in the observable range to be detectable.
In models where sterile neutrinos communicate with the SM via mixing with the active neutrinos,
the main decay channel of  keV sterile neutrino is $\nu_s \rightarrow \nu\nu\bar{\nu}$
with different combinations of flavors \cite{Pal:1981rm,Barger:1995ty}, which determines the lifetime of sterile neutrino.
From the requirement of the lifetime longer than $t_{\tiny {\rm Universe}}$, we get the bound
on the mixing angle $\theta$ between active and sterile neutrino \cite{Ade:2015xua,Boyarsky:2018tvu},
\begin{eqnarray}
\sin^2\theta \lesssim 3.3 \times 10^{-4} \left( \frac{10 {\rm keV}}{m_s}\right)^5.
\end{eqnarray}
A sub-dominant decay channel is the radiative decay of sterile neutrino into a photon and an active neutrino \cite{Lee:1977tib,Pal:1981rm,Barger:1995ty,Shrock:1982sc},
which occurs via 1-loop diagrams as shown in Fig. \ref{fig:nugamma}.
The decay width is proportional to $m^5_s \sin^2 2\theta$.
Since the active neutrino mass is negligible compared to the keV mass of the sterile neutrino,
the photon produced from the decay is monoenergetic.
Therefore, the smoking gun signature to search for the sterile neutrino DM is to observe
a monoenergetic X-ray at half of the sterile neutrino mass.

\begin{figure}
  \includegraphics[width=\linewidth]{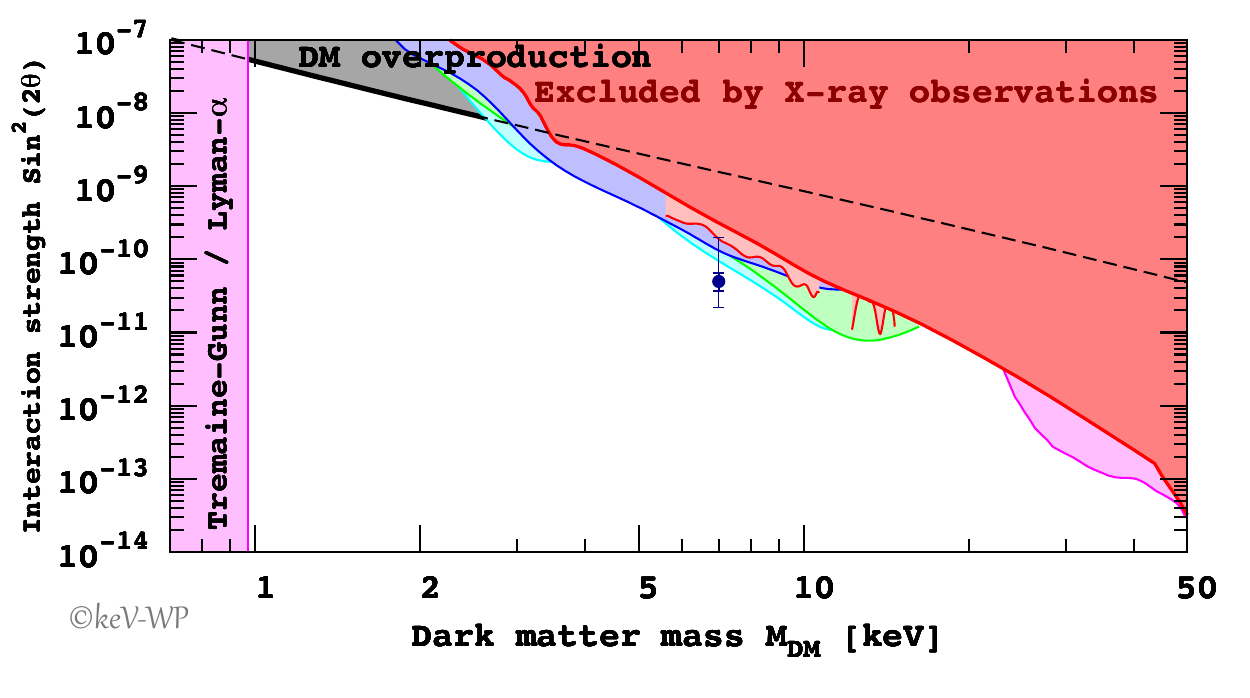} \caption{Bounds on  $\sin^2 2\theta$ as a function of sterile neutrino DM mass \cite{Adhikari:2016bei}.
}
\label{fig:exclusion_plot}
\end{figure}

There exist several types of the X-ray signal generated by the decay of sterile neutrino DM. 
First, such a decay can contribute to the diffuse X-ray background (XRB) throughout the histroy of the Universe\cite{Dolgov:2000ew,Abazajian:2001vt,Mapelli:2005hq,Boyarsky:2005us}.
The non-observation of the decay feature in the XRB signal leads us to a bound on the mixing angle $\theta$
and DM mass $m_s$\cite{Boyarsky:2005us,Abazajian:2006jc},
\begin{eqnarray}
\Omega_s \sin^2 2 \theta \lesssim 3\times 10^{-5} \left( \frac{1 {\rm keV}}{m_s}\right)^5.
\end{eqnarray}
There exists the total DM decay flux from the individual clusters of galaxies\cite{Abazajian:2001nj,Mapelli:2005hq}, which is of the same order
of that from the XRB but shows different spectral shape of an expected signal.
While the DM decay line in the XRB spectrum is broad, that from the cluster is narrow, whose width is determined by the spectral resolution of an X-ray detector or by the velocity dispersion of the DM halo of the cluster.
As shown in \cite{Boyarsky:2005us,Boyarsky:2006fg}, the signal in the narrow energy band centered on the line energy $E=m_s/2$ can be enhanced.
We can also observe the flux from the direction of individual dwarft spheroidal(dShp) galaxies \cite{Mateo:1998wg} 
which is expected to be comparable to the flux from individual galaxy clusters\cite{Boyarsky:2006fg,Boyarsky:2009rb,Boyarsky:2009af}.
Although DM mass concentrations nearby dSph galaxies have much smaller DM masses and their luminosity
in the DM decay line is lower than that from the galaxy clusters\cite{Mateo:1998wg},  the flux of the line gets increased due to
their proximity to the Milky Way \cite{Boyarsky:2006fg,Boyarsky:2009rb,Boyarsky:2009af}.
Finally, it is expected to observe the DM decay signal from the Milky Way halo whose strength is comparable
to that from the galaxy clusters, in spite of the much smaller DM mass in those structures.
Search for the DM decay signals in the keV-MeV mass range  from  the XRB, galaxy clusters, dShp galaxies and the Milky Way halo was conducted by using various X-ray telescopes (see Ref.\cite{Adhikari:2016bei}).
Non-observation of the decay line leads us to upper limits on the mixing angle $\theta$ as a function of
the sterile neutrino mass $m_s (\equiv M_{\rm DM})$, which are shown in Fig. \ref{fig:exclusion_plot} where
the colored regions are excluded.

\paragraph{Bounds From the Lyman-$\alpha$ Forest}:
The Lyman-$\alpha$ forest is a series of absorption lines shown in the spectra of galaxies and quasars
arising from the Lyman-$\alpha$ electron transition of the hydrogen atom \cite{Wyman:2013lza,Hamann:2013iba,Battye:2013xqa,Leistedt:2014sia,Palanque-Delabrouille:2014jca}.
It maintains cosmological information and could be used to probe the clustering of matter over a range of scales
from sub-Mpc up to few hundreds of Mpc and a range of
redshift from $z=2$ to $z=6$.
The absorption spectrum is a map of density fluctuations in the intervening intergalatic medium with peaks of density fluctuations at the density peaks of the absorbing gas\cite{Borde:2014xsa}.
Thus, the amplitude and shape of the power spectrum of  matter fluctuations measured through the Lyman-$\alpha$ forest observables in the spectra can be used to constrain cosmology.

The Sloan Digital Sky Survey (SDSS) \cite{York:2000gk} provided a much larger sample of 3035 medium-resolution quasar spectra  which allows the detailed measurements of 
the Lyman-$\alpha$ forest power spectrum \cite{McDonald:2004eu} so as to provide cosmological constraints \cite{McDonald:2004xn}.
A new measurement of the Lyman-$\alpha$ forest power spectrum in both 3-dimensonal and
1-dimensional redshift space has been implemented by using a much larger number of quasar spectra
at redshift $z>2$ that could be observed by SDSS III \cite{Eisenstein:2011sa}
through the Baryon Oscillation Spectroscopic Survey \cite{Dawson:2012va}.
While three-dimensional power spectrum uses only information from the flux correlation of pixel pairs in different quasar spectra leading to information on rather large scales, 
one-dimensional power spectrum uses the correlation of pixel pairs on the same quasar spectrum leading to information on smaller scales that are fundamental to constrain the physical parameters of the Lyman-$\alpha$ forest \cite{Borde:2014xsa}.

Lighter mass warm DM particles more easily escape gravitational potentials, and thus suppress structure on large scales, which can be constrained by the observed clustering on small scales of the Lyman-$\alpha$ forest
\cite{Markovic:2013iza}.
The structure formation is not directly sensitive to the properties of DM particles, but sensitive to their free streaming length $\lambda_{\rm FS}$.
A bound on $\lambda_{\rm FS}$ can be converted into a bound on the sterile neutrino DM mass if their momentum distribution is known.
For a thermal relic warm dark matter, $\lambda_{\rm FS}$ is given in a good approximation by
\cite{Bond:1980ha},
\begin{equation}
\lambda_{\rm FS}\sim 1 {\rm Mpc} \frac{\mbox{keV}}{m_{\rm DM}}
\frac{\langle p_{\rm DM}\rangle}{\langle p_{\nu}\rangle} ,
\end{equation}
where $\langle p_{\rm DM}\rangle$ is the average momentum of the DM particles and $\langle p_{\nu}\rangle \sim 1$ keV is the comoving momentum of active neutrinos when the sterile neutrino thermally produced becomes non-relativistic.
But, there is no such simple relation for the sterile neutrinos produced resonantly or from a heavy particle decay.

The most recent and comprehensive analysis of the Lyman-$\alpha$ forest from HIRES/MIKE for warm DM models has been made in \cite{Viel:2013apy}.
The conclusion is that the lower limit on $m_s$ is $3.3$ keV at 2$\sigma$ C.L. for a thermally produced  warm DM.
Similar analysis of the same high-resolution Lyman-$\alpha$ forest data had given constraints of $m_s>3.3$ keV\cite{Viel:2013apy} from HIRES/MIKE and $m_s>3.95$ keV from SDSSIII/BOSS \cite{Bhattacharyya:2017epv}. 
If one allows for a sudden jump in the temperature evolution the constraint above becomes weaker by about 1 keV.
Recent analysis  of the XQ-100 and HIRES/MIKE Lyman-$\alpha$ forest data sets\cite{Irsic:2017ixq} provides new constraints on the free-streaming of warm DM, which is given in term of
the mass of a thremally produced  warm DM particle by $m_s > 5.3$ keV at $2\sigma$.
These bounds rely on the assumption that the DM momentum distribution is not too different from the thermal spectrum.

\subsection{Light Sterile Neutrino As a Mediator for Leptogenesis} \label{low-scale-lepto}
Here, we shortly show how light sterile neutrino plays a crucial role in successfully achieving low scale leptogenesis.
A concrete model will be provided later, where  both right amount of relic abundance of sterile neutrino and low scale leptogenesis
can be simultaneously achieved.
On top of type I seesaw model, introducing an extra light sterile neutrino $\chi$ and a singlet scalar field $\phi$  and allowing new Yukawa interaction,
$ \bar{\chi}\phi N_{R_i}$ with heavy right-handed neutrino $N_{R_i}$, we can achieve low scale leptogenesis.
The coupling constant of this interaction can be in general complex which is the source of CP violation required.
In this scenario, leptogenesis is realized by the decay of lightest heavy Majorana neutrino $N_{R_1}$ into the Higgs boson and leptons in the SM
 before the scalar fields get vacuum expectation values.
The CP violation required for leptogenesis is given by
\begin{eqnarray}
\epsilon_{1} &=& -\sum_i\left[\frac{\Gamma(N_{R_1}
\to \bar{l_i}H^{\ast}) - \Gamma(N_{R_1} \to l_i H) }{\Gamma_{\rm
tot}(N_{R_1})}\right] , \label{epsilon}
\end{eqnarray}
with  $\Gamma_{\rm tot}(N_{R_1})$ being
 the total decay rate of $N_{R_1}$.
\begin{figure}[h!]
\begin{center}
\includegraphics[width=3.2in]{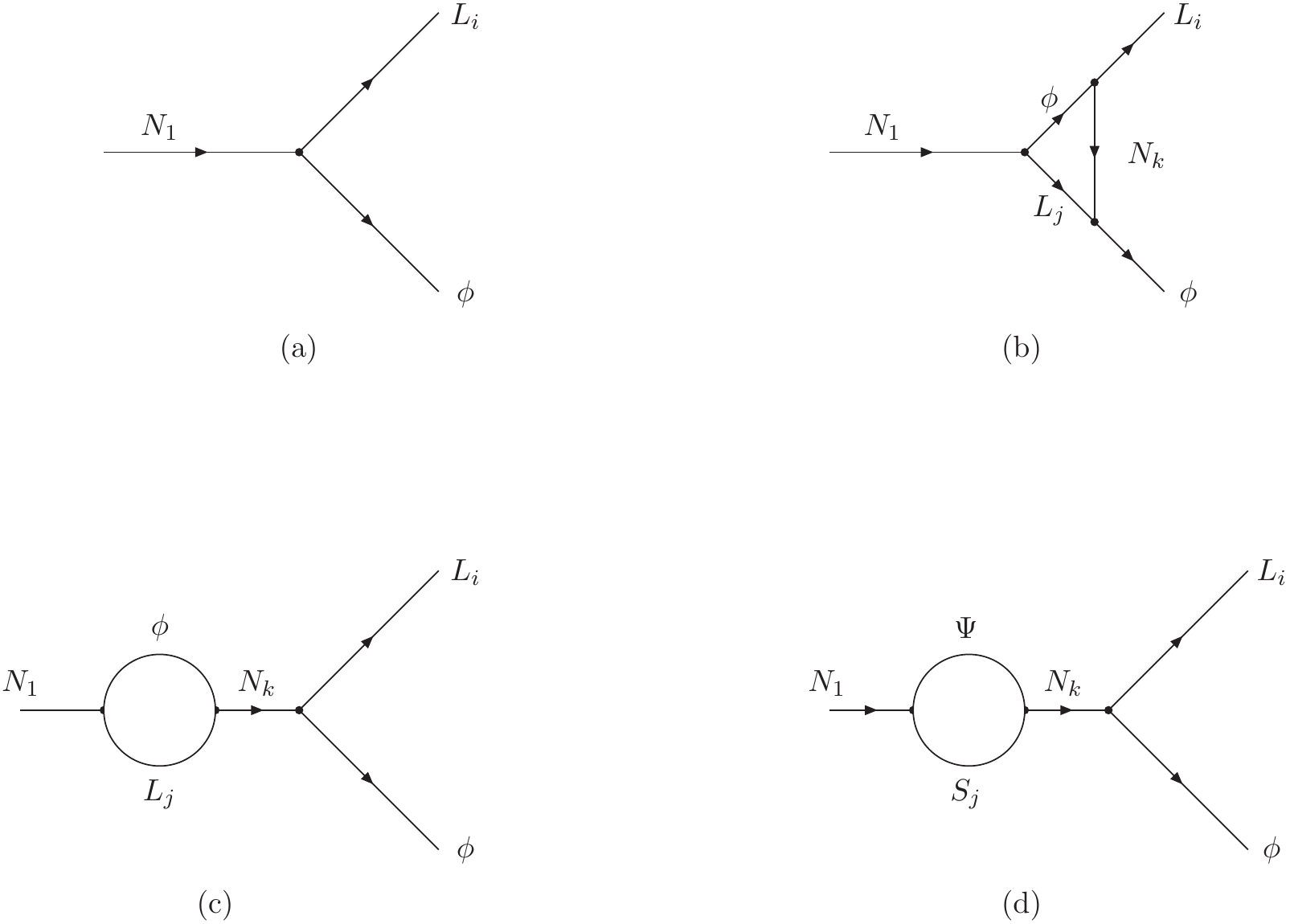}
\caption{Diagrams contributing to lepton asymmetry \cite{Kang:2006sn}.}
\label{fig:low-lepto}
\end{center}
\end{figure}
Fig. \ref{fig:low-lepto} shows the structure of the diagrams contributing to
$\epsilon_{1}$.
In this model, 
 on top of the diagrams for the standard
leptogensis \cite{Covi:1996wh}, a new vertex diagram arising due to the Yukawa interaction $Y_{\chi} \bar{\chi}\phi N$ contributes to $\epsilon_1$.
The key idea on the realization of low scale leptogenesis is the enhancement of $\epsilon_1$ due to the new vertex contribution to $\epsilon_1$.
The details on this mechanism for low scale leptogenesis will be provided in next subsection.

%
\subsection{A Model for keV Sterile Neutrino DM
}
%
%
In this section, we review on a renomalizable model with keV sterile neutrino
to show how the sterile neutrino can be a good DM candidate and 
play an important role in low scale leptogenesis.
In this model, the relic abundance is related with lepton asymmetry required for successful baryogenesis through leptogenesis.
As will be shown, tiny neutrino masses can be achieved via  so-called double seesaw mechanism
in the model.
\paragraph{Setup} :
For our purpose, we extend the type I seesaw model by introducing a singlet neutrino, $\chi$, and a singlet scalar field, $\phi$.
In this model, we consider the possibility that the sterile neutrino  $\chi$ is a keV scale DM.
The  Yukawa interactions and mass terms associated with neutrinos  in the Lagrangian we consider are given by \cite{Kang:2006sn,SungCheon:2007nw,Kang:2014mea},
\be
{\cal L}=M_{N_i}N_{R_i}^T N_{R_i}+Y_{D_{ij}} \bar{\L}_i
H  N_{R_j}+ Y_{\chi_{i}} \bar{\chi}\phi N_i -\mu \chi^T \chi +h.c.~,
\label{Lag}
\ee
where $\L_i,N_{R_i}$, and $ H $ stand for an SU(2)$_L$ lepton doublet with flavor index $i$, a right-handed singlet neutrino, and an SU(2)$_L$ scalar doublet field, respectively, and we take the charged lepton diagonal basis.
We see that the Lagrangian Eq.(\ref{Lag}) has a $Z_2$ symmetry under which the fields $\phi$ and $\chi$ are odd, and all the other fields are even.
This $Z_2$ symmetry, if unbroken,  makes the lightest singlet neutrino
$\chi$ stable and thus be a DM candidate.
However, the $Z_2$ symmetry is broken once the singlet scalar field $\phi$ takes a non-zero vacuum expectation value (VEV). Although $\chi$ is no longer stable in that case, it is still a viable DM if its lifetime is much longer than $t_{\tiny {\rm Universe}}$, as will be shown later. 
The scalar potential is given by \cite{Krasnikov:1997nh,Babu:2014pxa} 
\be
{\cal {L}}_{Scalar}=\frac{\mu_H^2}{2} H^{\dagger} H+\frac{\lambda_H}{4}(H^{\dagger}H)^2+\frac{\mu_\phi^2}{2} \phi^2 +\frac{\lambda_\phi}{4}\phi^4+\frac{\lambda_{H \phi}}{2}(H^{\dagger}H)\phi^2.
\ee
Defining VEVs of the neutral components of the $H$ and the $\phi$ fields as $v$ and $v_{\phi}$, respectively, we expand the fields as $H \rightarrow \frac{h}{\sqrt{2}}+v$ and $\phi \rightarrow \phi+v_{\phi}$. This gives us a $2\times2$ mass squared matrix for the scalar fields  as follows,
\be
\begin{pmatrix} \lambda_H v^2&\sqrt{2}\lambda_{H\phi}v_{\phi}v\\\sqrt{2}\lambda_{H\phi}v_{\phi}v&2\lambda_\phi v_{\phi}^2 \end{pmatrix}.
\ee
Assuming $\lambda_{H\phi} << \lambda_H,\lambda_\phi$, the eigenvalues of this mass squared matrix are given by
\begin{eqnarray}
m_{H^{\prime}}^2 &=& \lambda_H v^2+\frac{\lambda_{H\phi}v v_{\phi}}{\sqrt{2}}\tan{2\theta}, \notag \\
m_{\phi^{\prime}}^2 &=& 2\lambda_\phi v_{\phi}^2-\frac{\lambda_{H\phi}v_{\phi} v}{\sqrt{2}}\tan{2\theta},
\label{eq:scalar}
\end{eqnarray}
where $\theta$ is the mixing angle between $H$ and $\phi$ explicitly given by
\be
\tan{2\theta}=\frac{2\sqrt{2}\lambda_{H\phi}v_{\phi}v}{\lambda_Hv^2-2\lambda_\phi v_{\phi}^2}. 
\ee

We take  $\lambda_{H\phi} > 10^{-6}$ ~~\cite{Kusenko:2006rh,Petraki:2007gq} in order to make 
$\phi$ be an appropriate candidate of the heat bath.
 On the other hand, we demand that the mixing angle $\theta$ is small enough to not affect the SM like Higgs phenomenology. Thus, we take $\lambda_{H \phi} =5 \times 10^{-5}$ in the numerical analysis. For such a small $\lambda_{H\phi}$, we can ignore its contributions to the scalar masses given in Eq.~({\ref{eq:scalar}}), as well as its effects on the LHC Higgs boson signals. 
Then, we may take $m_H=125$ GeV which is achieved for $v=246$ GeV and $\lambda_H=0.26$. 
We also choose $v_{\phi}=100$ GeV and $\lambda_\phi=0.5$, which gives rise to $m_\phi=100$ GeV which will be taken as a benchmark point for achieving the right DM abundance.
%
%
\paragraph{Light Neutrino Masses}:
From the Lagrangian Eq.(\ref{Lag}), we see that
the neutrino mass matrix in the basis $(\nu_j,N_{R_i},\chi)$ is given as \cite{Kang:2006sn,SungCheon:2007nw,Kang:2014mea}
\be
M_{\nu}=\left(\begin{array}{ccc}
 0 & m_{D_{ij}} & 0 \\
 m_{D_{ij}} & M_{N_{ii}} & M_{\chi_{i}} \\
 0 & M_{\chi_{i}} & -\mu \end{array}\right), \label{massmatrix}
\ee
where $m_{D_{ij}}=Y_{D_{ij}}\langle H \rangle$ and $M_{\chi_{i}}=Y_{\chi_{i}}\langle \phi \rangle$. Assuming that $M_N \gg M_{\chi} \gg \mu, m_{D}$,
 one can get the light active neutrino mass matrix and  their mixing with $\chi$  are given by
\be
m_{\nu} &\simeq & \frac{1}{2}\frac{m_D}{M_{\chi}}~\mu~ \left(\frac{m_D}{M_{\chi}}\right)^T,
                  \label{dsw}\\
\tan2\theta_{\chi} &=& \frac{2m_D M_\chi}{M^2_\chi+4\mu M_N-m_D^2}~, \label{mixing}
\ee
where we have omitted the indices of the mass matrices, $m_D,M_\chi,M_N$ and $\mu$,
for simplicity.

On the other hand, the  mass of sterile neutrino, $\chi$,  is approximately
given by
\begin{eqnarray}
m_{\chi} \simeq \mu + \frac{M^2_\chi}{4M_N}. \label{sterile}
\end{eqnarray}
Further assuming that $M^2_\chi \ll 4 \mu M_N$, 
$\theta_\chi$ and $m_{\chi}$ approximately become
\be
\tan2\theta_{\chi} &\simeq & \sin2\theta_{\chi}  \simeq 
  \frac{m_D M_{\chi}}{2\mu M_N},  \label{mixing1}\\
m_{\chi} &\simeq &\mu .
 \label{sterile2}
\ee
Taking $\mu \sim 7.1$ keV, the sterile neutrino is regarded as DM leading to 3.55-keV
peak in the galactic X-ray spectrum.
Then, we see from Eq.(\ref{dsw}) that $m_D/M_\chi \simeq 3.8 (1.7)\times 10^{-3}$ for $m_{\nu}\simeq 0.05 (0.01)$ eV, corresponding to the atmospheric (solar) neutrino mass scale in the hierarchical spectrum.
If low scale leptogenesis is to be achieved,  the lightest $M_N$ should be taken to be a few TeV.
Once we take $M_N\sim 10$ TeV,  we are led from Eq. (\ref{mixing}) to
\be
\theta_{\chi}  \simeq \frac{m_D M_\chi}{4 \mu M_N}\sim \left(\frac{M_\chi}{0.86 (1.3)\times 10^{10} \mbox{eV}}\right)^2. \label{s-mixing}
\ee
In order for the sterile neutrino to be DM with a mass of $7.1$ keV, a requirement is that
$\sin2\theta_{\chi}\sim 10^{-5}$ \cite{Bulbul:2014sua,Boyarsky:2014jta}. 
Combining this constraint with Eq.(\ref{s-mixing}), we find the scale of $M_{\chi}$ to be
around $ 20(30)$ MeV and, thus,  $m_{D}$ to be 70(50) keV.
Those scales of $m_{D}$  and  $m_{\chi}$ are achieved by taking $Y_{D}\sim 10^{-6}$ and
$Y_{\chi}\sim 10^{-4}$, respectively for $\langle H \rangle\simeq 246$ GeV and $\langle \phi \rangle \sim 100$ GeV.
When we take $M_N$ to be less than 10 TeV, then we get a smaller $M_\chi$ and $m_D$.

We note that such a small mixing angle $\theta_\chi$ ensures that
sterile neutrinos $\chi$ were never in thermal equilibrium in the early
Universe and thus their abundance must be smaller than that in thermal equilibrium.
Let us show how the correct abundance of the sterile neutrino
can be achieved via freeze-in decay of the scalar field $\phi$.

%
\paragraph{Production of DM}:
%
Now, let us study how the relic abundance of $\chi$ with a mass of keV scale  can be computed.
We note that the interactions with couplings $Y_{\chi_1}$ were never in
thermal equilibrium in early Universe because  it is so small that tiny neutrino masses, as well as small  mixing angle $\theta_{\chi}$ could be realized.
It is worthwhile to note that  a large value of $Y_{\chi_2}$ is essential in  achieving not only low scale leptogenesis but also the correct amount of relic abundance.
In our model, the production of $\chi$ through $H+\nu \rightarrow \phi +\chi$ is turned out to be  negligibly small compared to their production through the decay of  the scale field $\phi$. 
So, we focus on the production of $\chi$ via so-called in-equilibrium  decay of  $\phi$  \cite{Kusenko:2006rh,Petraki:2007gq,Merle:2015oja}. 

After the heavy Majorana neutrino $N_{2(3)}$ decouples and $\phi$ gets VEV, an effective interaction term, $|Y_{\chi_{2}}|^2 (v_{\phi} /M_{N_2}) \phi \bar{\chi}\chi $, is generated, through which 
$\phi$ decays into a pair of $\chi$.
Thanks to rather large value of $\lambda_{H\phi}$, $\phi$ is in equilibrium until it decouples at 
the freeze-out temperature $T_f$.
The rate for the decay process, $\phi \rightarrow \chi \chi $, is given by
\begin{eqnarray}
\Gamma(\phi\rightarrow \chi \chi) \simeq \frac{|Y_{\chi_2}|^4}{64\pi}\frac{v_{\phi}^2}{M_{N_2}^2}m_{\phi}. \label{decay-eta}
\end{eqnarray}
In general, this decay process takes place through both  in-equilibrium and out-of-equilibrium decay of $\phi$ into a pair of $\chi$ \cite{Kusenko:2006rh,Petraki:2007gq,Merle:2015oja}. The equilibrium production occurs at temperature above $T_f$ while at temperature below $T_f$ the production of DM continues to occur via the out-of-equilibrium decay of $\phi$. 
In this work, we will choose a parameter space where the decay process takes place through in-equilibrium  decay of $\phi$ into a pair of $\chi$
at temperatures $T>T_f$ \cite{Kusenko:2006rh,Petraki:2007gq,Merle:2015oja}.
\footnote{From our numerical analysis, it turns out that a 7-keV sterile neutrino DM is not realized
via out-of-equilibrium decay of $\phi$ mainly because the production mechanism leads to very large free streaming length that would be ruled out by the Lyman-$\alpha$ bound \cite{Viel:2005qj,Seljak:2006qw,Viel:2006kd}.}.
The freeze-out temperature of $\phi$, $T_f$, is estimated from the usual criterion given as
\begin{equation}
\left. \frac{\left< \sigma v_{\rm rel} \right> n_{\rm eq}}{H} \right|_{T=T_f} \simeq 1,
\label{criterion}
\end{equation}
where $v_{\rm rel}$ is the velocity, $n_{\rm eq}$ is the equilibrium distribution, which is assumed to be Maxwell-Boltzmann in this case, $H$ is the Hubble constant, and $\sigma$ is the annihilation cross-section \cite{McDonald:1993ex}.
Solving Eq.(\ref{criterion}) for a new parameter $r_f = \frac{m_\phi}{T_f}$  results in $r_f \simeq 13$.

The distribution function of sterile neutrino $f_{\chi}$  can be obtained by solving the following Boltzmann equation:
\be
\frac{\partial f_\chi}{\partial t}- H p\frac{\partial f_\chi}{\partial p} = C[\phi \rightarrow \chi\chi]
+C[\nu_a \leftrightarrow \chi], \label{be1}
\ee
where
\be
C[\phi \rightarrow \chi\chi] &&\equiv  
\frac{2 m_{\phi} \Gamma(\phi\rightarrow \chi \chi)}{p^2}
\int^{\infty}_{p+\frac{m^2_{\phi}}{4p}}  \frac{dE}{\exp(E/T)-1}
, \nonumber \\
C[\nu_a \leftrightarrow \chi] &&\equiv h_{\alpha}(p,T) [f_{\nu_{\alpha}}-f_\chi].
\label{C-functions}
\ee
The $C[\phi \rightarrow \chi\chi]$ corresponds to the contribution arising from
the decay of $\phi$ into a pair of $\chi$, and
the $C[\nu_a \leftrightarrow \chi]$  to that due to the active-sterile mixing.
Here,
\be
h_{\alpha}(p,T)&&=\frac{1}{8}\frac{\Gamma_{\alpha}(p,T)\Delta^2(p) \sin^2 2\theta}
{\Delta^2(p) \sin^2 2\theta + D^2(p,T)+[\Delta(p) \cos 2\theta -V(p,T)]^2}, \nonumber \\
\Gamma_{\alpha}&&\simeq c_{\alpha} G^2_F ~p ~ T^4,  ~~~~
D(p,T)=\frac{1}{2}\Gamma_{\alpha}, ~~~~
\Delta(p)\simeq \frac{m^2_{\chi}}{2p}, \nonumber \\
f_{\nu_{\alpha}}&&=\frac{1}{\exp(E_{\nu_\alpha}/T)+1},
\ee
where $\alpha=e, \mu, \tau$, $\theta$ denotes the active-sterile mixing angle, and  $G_F \approx 1.166 \times 10^{-11}$ $\text{MeV}^{-2}$ is the Fermi constant. 
We take $c_{\alpha} $ to be 1.27 (0.92) for $\alpha=e (\mu, \tau)$, and 
the explicit form of $V(p,T)$ is presented in Ref. \cite{Abazajian:2001nj}.
Changing to the variables $r=\frac{m_{\phi}}{T}$ and $x=\frac{p}{T}$,
Eq. (\ref{be1}) is rewritten as follows:
\be
\frac{\partial f_\chi (r,x)}{\partial r}=\left( \frac{r M_0}{m^2_{\phi}}\right) \left[
\frac{2r\Gamma_{\phi}}{x^2} \int^{\infty}_{z_{\rm min}} \frac{dz}{\exp(z)-1}
+  h_{\alpha}(r,x)\left( f_{\nu_{\alpha}}(r,x)-f_\chi (r,x)\right) \right],
\label{bef}
\ee
where $M_0 = \sqrt{\frac{45 M_{Pl}^2}{4 \pi^3 g^\ast_{\rho}}}$ is the reduced Planck mass,$z_{\rm min}=x+\frac{r^2}{4x}$ and $\Gamma_{\phi}=\Gamma(\phi \rightarrow \chi \chi)$.
In the analysis, we see that the first term on the right-hand side of Eq. (\ref{bef}) is
dominant for $ T \geq T_f$, whereas $\chi$ is mainly produced via oscillations between the active and the sterile neutrinos for $T<T_f$.
Thus, solving Eq. (\ref{be1}) in the limit of constant $g^{\ast}_{\rho} $ by turning off the second term for $ T \geq T_f$ and the first term for  
$T< T_f$, one can obtain the approximated form of $f_{\chi}(r,x)$  given by
\be
f_\chi (r,x) \simeq &&\exp\left(\int^{r}_{r_f} h_s(r^{\prime},x)dr^{\prime}\right)
\left[f_{\chi}^{in}(r_f,x)
\right. \nonumber \\
&& \left. +\int^{r}_{r_f} dr^{\prime}  \exp\left(-
\int^{r^{\prime}}_{r_f} h_s(r^{\prime \prime},x) dr^{\prime \prime}\right)
\frac{h_s(r^{\prime},x)}{\exp(x)+1}
 \right], \label{bes1}
\ee
where 
\be
&&h_s(r,x)\equiv -\frac{r M_0}{m^2_{\phi}} h(r,x), \nonumber \\
&& f^{in}_{\chi}(r_f,x)=\frac{2M_0\Gamma}{3 m^2_{\phi}} \left[ \frac{r_f^3}{x^2}\ln (1-e^{-x-\frac{r_f^2}{4x^2}})^{-1}+8 x^2 \int^{1+\frac{r_f^2}{4x^2}}_1 \frac{(z-1)^{3/2}dz}{e^{xz}-1}\right].
\ee 
Because the production of the sterile neutrino via the decay of singlet scalar field mostly happens around $T_{\rm{prod}}=m_{\phi}/2.3$ \cite{Petraki:2007gq},  we take $g^{\ast}_{\rho}(T_{\rm{prod}})=88$  in the calculation of $f^{in}_{\chi}(r_f,x)$ whereas  $g^{\ast}_{\rho}(T)$ in $h_s(r,x)$ is taken to be 63.5 because the effect of oscillation exhibits a peak at around $ 100~ \rm{MeV} \leq T \leq 1~ \rm{GeV}$.
We found from numerical analysis that the constribution of the second term on the right-hand side of Eq. (\ref{bes1}) to $f_{\chi}(r,x)$ is less than  $1\%$ \cite{Kang:2014mea}.

Using  Eq.(\ref{bes1}), the relic abundance presented in terms of
the yield $Y_{\chi}=n_{\chi}/S$,  defined by the ratio of the number density $n_s$ to the entropy $S (=2\pi^2 g_s^{\ast}T^3/45)$, is given by
\be
Y_\chi (r)=\frac{45}{4\pi^4 g_s^{\ast}}\int^{\infty}_{0}  x^2 f_\chi (r,x) dx.
\ee
The final result of the relic density of the sterile neutrino is written as
\begin{eqnarray}
\Omega_{\chi} h^2 &=& 2.733 \times 10^8 \left(\frac{Y_{\infty}}{\xi_d}\right) \left(\frac{m_{\chi}}{\rm{GeV}}\right),
\end{eqnarray}
where $Y_{\infty}\equiv Y(r\rightarrow \infty)$ and $\xi_d=g^{\ast}_s(T_{\rm{prod}})/g^{\ast}_s(0.1 \rm{MeV}))\simeq 27$ is the dilution factor \cite{Petraki:2007gq}.
The DM relic density observed by PLANK  \cite{Ade:2013zuv} is 
$ \Omega_{\rm{DM}} h^2=0.1199\pm 0.0027$, and we get  its central value
by taking $m_{\phi}=v_{\phi}=100$ GeV,  $M_{N_2}=10$ TeV, and
$Y_{\chi_2}=2.8 \times 10^{-3}$ in this model.
This indicates that  the required amount of lepton asymmetry and the correct relic density 
 \cite{Ade:2013zuv}  can be achieved at a relatively low seesaw scale (a few TeV scale) in this model.

To see how warm the sterile neutrino DM in the scenario we consider is, we need to calculate the free-streaming length.
The average momentum of the sterile neutrino is calculated with the help of the relation
\begin{equation}
\frac{\left< p \right>}{T} =\frac{ \int_0^\infty\frac{d^3 p}{(2 \pi)^3} p f_\chi (p,T)}{ T \int_0^\infty\frac{d^3 p}{(2 \pi)^3} f_\chi (p,T)} = \frac{ \int_0^\infty dx~ x^3  f_\chi (r,x)}{ \int_0^\infty dx~x^2 f_\chi (r,x)}.
\end{equation}
Plugging an asymptotic ($r \rightarrow \infty $) form of $f_{\chi}$ into the above equation, we get the average momentum given by
\begin{equation}
\frac{\left< p \right>}{T} = \frac{2.477}{\xi^{1/3}_d},
\label{avemom}
\end{equation}
where the momentum $\left<p\right>$ is red-shifted by a factor $\xi^{1/3}_d$  as shown in\cite{Petraki:2007gq,Shaposhnikov:2006xi}.
This result indicates that the sterile neutrino DM is colder than the thermally produced warm DM.

Using the approximated form of the free-streaming length \cite{Petraki:2007gq,Bond:1980ha},
$\lambda_{FS} \approx 1.2~{\text{Mpc}} \left(\frac{\text{keV}}{m_\chi}\right) \left( \frac{\left<p\right>}{3.15 T} \right)$,
we get
$\lambda_{FS} \approx 0.044~ \text{Mpc}$ for the sterile neutrino DM.
This result is consistent with the constraint coming from the observation of
the Lyman-$\alpha$ forest which requires it to be less than 0.11 Mpc \cite{Viel:2005qj,Seljak:2006qw,Viel:2006kd}.

%
%
\paragraph{Low Scale Leptogenesis}:
In section \ref{low-scale-lepto}, we have explained how the light sterile neutrino can play a crucial role in achieveing 
low scale leptogenesis.
Here, we study the details on how the mechanism for low scale leptpogenesis works.
Calculating $\epsilon_{1}$ given in Eq.(\ref{epsilon}) based on the diagrams shown in Fig. \ref{fig:low-lepto}, we get the explicit form of  CP asymmetry given by
\begin{equation}
\epsilon_1 = \frac{1}{8\pi} \sum_{k\ne 1} \left[ ( g_V(x_k)+
g_S(x_k)){\cal V}_{k1} + g_S(x_k){\cal S}_{k1}\right],
\end{equation}
where $g_V(x)=\sqrt{x}\{1-(1+x) {\rm ln}[(1+x)/x]\}$,
$g_S(x)=\sqrt{x_k}/(1-x_k) $ with $x_k=M_{N_k}^2/M_{N_1}^2$ for
$k\ne 1$,
\begin{equation}
  \label{eq:epsilon}
{\cal V}_{k1}={{\rm Im}[(Y_D Y_D^\dagger)_{k1}^2]
 \over (Y_D^\dagger Y_D)_{11} +|Y_{\chi_1}|^2},
\end{equation}
and
\begin{equation}
{\cal S}_{k1}={{\rm Im}[(Y_D Y_D^\dagger)_{k1}(Y_{\chi_k}
Y_{\chi_1}^\dagger)]
 \over (Y_D^\dagger Y_D)_{11} +|Y_{\chi_1}|^2}.
\end{equation}
In the calculation,  we have used $\Gamma_{\rm tot}(N_{R_1})$ given to leading order  by
\begin{equation}
  \label{eq:vv}
\Gamma_{\rm tot}(N_{R_1})={(Y_D^\dagger Y_D)_{11}+|Y_{\chi_1}|^2
\over 4\pi}M_{N_{1}}.
\end{equation}

As shown in Ref. \cite{Kang:2006sn,SungCheon:2007nw},  the new contribution to $\epsilon_1$ becomes important for the case of  $M_{N_1}\simeq M_{N_2} \ll M_{N_3}$, resulting in
 \begin{eqnarray}
 \epsilon_1 & \simeq & -\frac{1}{16\pi}
           \frac{M_{N_2}}{v^2}\left[\frac{Im[(Y^{\ast}_D m_{\nu}Y^{\dagger}_D)_{11}]}
           {(Y_D^\dagger Y_D)_{11}+|Y_{\chi_1}|^2}
        +\frac{ Im[(Y_DY^{\dagger}_D)_{21}(Y_{\chi_2}Y_{\chi_1}^{\dagger})]}
                   {(Y_D^\dagger Y_D)_{11}+|Y_{\chi_1}|^2}\right]R~,
                   \label{epsilon2}
 \end{eqnarray}
where $R$ is a resonance factor defined by $R \equiv
|M_{N_1}|/(|M_{N_2}|-|M_{N_1}|)$.
Imposing the
out-of-equilibrium condition, $\Gamma_{N_1} < H|_{T=M_{N_1}}$ with
the Hubble expansion rate $H$, we get the upper
bound on  $Y_{\chi_1}$ ,
\begin{eqnarray}
\sqrt{\sum_i|Y_{\chi_1}|^2}<3\times
10^{-4}\sqrt{M_{N_1}/10^9(\mbox{GeV})}.
\end{eqnarray}
However,  since $Y_{\chi_2}$ is not constrained by the
out-of-equilibrium condition,  it can be so large that $\epsilon_1$ could be enhanced
without the enhancement of $R$ by tiny splitting between $M_{N_1}$ and $M_{N_2}$.
Taking $Y_{\chi_2}=\kappa_i (Y_D)_{2i}$ with constant $\kappa_i$ and $(Y_D)_{1i}\sim Y_{\chi_1}$, the upper limit of the second term of Eq. (\ref{epsilon2}) is given by $\kappa_i M_{N_2}\sqrt{\Delta m_{atm}^2}R/16\pi v^2$.

The generated B-L asymmetry is calculated through the relation, $Y_{\rm B-L}=-\xi
\epsilon_1 Y^{\rm eq}_{N_{R_1}}$, where $\xi$ is the efficiency factor and 
$Y^{\rm eq}_{N_{R_1}}\simeq
\frac{45}{\pi^4}\frac{\zeta(3)}{g_{\ast}k_B} \frac{3}{4}$ with the
Boltzmann constant $k_B$.
 In this model, a new process, $\chi \phi \rightarrow lH^0$,  mediated by $N_{R_{2(3)}}$,  dominantly
contributes to $\xi$.
To a good approximation, $\xi$ can be easily estimated by replacing
$M_{N_1}$ in the case of the canonical seesaw model with $M_{N_1} (Y_{\chi_2}/(Y_D)_{2i})^2$ \cite{Buchmuller:2002rq}.
From our numerical analysis, it turns out that successful leptogenesis can be achieved
for $M_{N_1}\sim 10^{4}$ GeV, provided that
$\kappa_i=Y_{\chi_2}/(Y_D)^{\ast}_{2i}\sim 10^{3}$ and $\delta M_N (\equiv M_{N_2}-
M_{N_1})\sim O (\mbox{GeV})$, avoiding the severe fine-tuning for the mass difference between $M_{N_1}$ and $M_{N_2}$ required for the resonant leptogenesis \cite{Pilaftsis:1997jf}. 

\begin{figure}[h!]
\begin{center}
\includegraphics[width=3.2in]{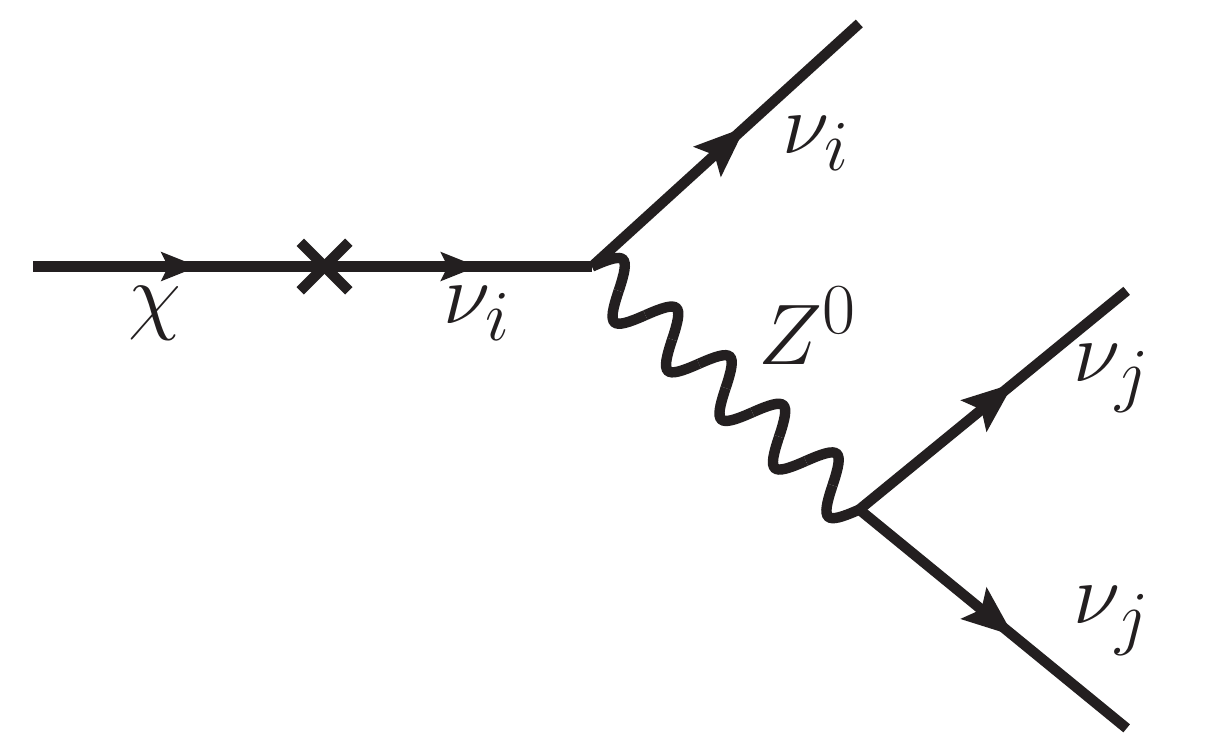}
\caption{Sterile neutrino decay into three active neutrinos \cite{Kang:2014mea}.}
\label{fig:3nu}
\end{center}
\end{figure}

%
\paragraph{Decay Signature of DM}:
%
%
The $Z_2$ symmetry introduced in the model is broken when $\phi$ gets a VEV. As a result, 
the sterile neutrino DM can not be stable any longer and decay at late time.
The primary decay channel  is $\chi\rightarrow \nu\nu\bar{\nu}$ as shown in Fig.~{\ref{fig:3nu}}. The decay width for this decay mode is given as
\begin{eqnarray}
\Gamma(\chi\rightarrow \nu\nu\bar{\nu}) &\simeq & G_F^2 \sin^2 {2\theta_{\chi}} \left( \frac{m_\chi^5}{768 \pi^3}\right) \notag \\
&\simeq &8.7 \times 10^{-31}~\text{sec}^{-1}\left(\frac{\sin^2 {2\theta_\chi}}{10^{-10}}\right) \left(\frac{m_\chi}{1~\text{keV}}\right)^5.
\end{eqnarray}
As  shown in Fig. \ref{fig:nugamma}, the sterile neutrino DM can also radiatively decay into an active neutrino along with the emission of a photon.
The decay width for the radiative processes  of the sterile neutrino DM is given by
\begin{equation}
\Gamma(\chi\rightarrow \nu\gamma) \simeq 6.8\times10^{-33}~\text{sec}^{-1} \left(\frac{\sin^2 {2\theta_\chi}}{10^{-10}}\right) \left( \frac{m_\chi}{1~\text{keV}}\right)^5.
\end{equation} 
In fact, the decay width for  the radiative processes  of the sterile neutrino DM is inversely proportional to the charged lepton mass inside the loop , and thus an electron results in a much stronger limit on the lifetime of the sterile neutrino DM. Thanks to the smallness of the electron Yukawa coupling,
 the contribution from the first pair of diagrams far exceed that from the other involving Yukawa couplings. 
The second pair of diagrams is suppressed by a factor of $\left( \frac{Y_e}{g} \right)^2$ $\sim$ $10^{-11}$ with the electron Yukawa coupling $Y_e$ and the $SU(2)_L$ gauge coupling $g$.
Therefore, only the first pair of diagrams places a meaningful limit on the lifetime of the sterile neutrino DM .

\paragraph{3.5 keV Line:}
The detection of an unidentified X-ray line at $ E \simeq 3.5$ keV has been reported
in the stacked spectrum of galaxy clusters,  in the Andromeda galaxy and many other galaxies including the Perseus galaxy \cite{Bulbul:2014sua,Boyarsky:2014jta}.
Although there exist some difficulties in interpreting the origin of the emission line due to inherent uncertainty
in the astrophysical backgrounds and systematic errors in the calibration, any interpretation of the emission line  in terms of  decaying DM is plausible if it is safe from all the constraints.
If the mass of the sterile neutrino DM is 7.1 keV, the emitted photon will have an energy of 3.55 keV, which
can be responsible for the observed unidentified 3.5 keV X-ray line.
In our model, the sterile neutrino $\chi$ with a mass of 7.1 keV has the decay width 
, $\Gamma_{total}\sim~10^{-26}~\text{s}^{-1}$, for  $\sin^2 {2\theta_\chi}\sim 10^{-10}$.
This result indicates that  $\chi$ has a lifetime much larger than the age of the universe, and thus
can be a good candidate for DM while being responsible for the 3.5 keV X-ray line.
%

\section{Conclusion and Prospect}
Some anomalous results from SBL neutrino experiments searching for neutrino oscillations are indicative of 
the existence of light sterile neutrinos.
In this review, we have discussed how the neutrino anomalies can be explained in terms of neutrino oscillations
with $\Delta m^2\sim O(1)$ ${\rm eV}^2$ and $\sin^2 2\theta\sim$ a few $\times 10^{-2}$ in the 3+1 model.
The excess of  $\bar{\nu}_e$  flux at LSND is in agreement with the observation of 
electron neutrino excess in both neutrino and antineutrino modes at MiniBooNE experiment.
Combining those two results leads to a global significance exceeding $6 \sigma$, and no-oscillation hypothesis for all $\protect\nua{e}$
appearance data is shown to be disfavored compared with the best fit at  $\Delta m^2_{41}\sim 0.6~{\rm eV}^2$ and $\sin^2 2\theta_{\mu e}\sim 0.04$.
The simultaneous explanation of both reactor antineutrino and Gallium anomalies requires values of $\Delta m^2 \sim 1.7~ \mbox{eV}^2$ and $\sin^2 2\theta \sim 0.1$ ,  and  the inclusion of the data from NEOS tends to shift downward the best fit value of the mixing angle to $\sin^2 2\theta  \sim 0.08$.
While both combinations of $\protect\nua{e}$ appearance and disappearance experiments show a robust indication
in favor of sterile neutrino,  no obvious anomaly has not been observed in the  $\protect\nua{\mu}$ disappearance channel.
As a result, the global fit to SBL data leads to a severe appearance-disappearance tension that should be resolved in future neutrino experiments.
Several new reactor neutrino experiments would probe the $\protect\nua{e}$ appearance channel with improved precision,
measuring the energy spectrum at different distances so as to obtain oscillation results that are free from the neutrino flux calculation.

The impacts of light sterile neutrinos in cosmology, particularly, taking into account two  intriguing cases of sterile neutrinos having eV or  keV scale  masses have been addressed.
It has been shown how the properties of the light sterile neutrinos such as a mass and mixing with active neutrinos can be constrained by several cosmological observations such as BBN, CMB , LSS and so on.
The keV sterile neutrino is a well motivated DM candidate. We have reviewed several mechanisms of how such a neutrino can be produced in the early Universe and
how its properties can be constrained by several cosmological observations.
We have also discussed the possibility that  keV-scale sterile neutrino can be a good DM candidate and  play a key role in achieving low scale leptogenesis simultaneously by introducing a model where an extra light sterile neutrino is added on top of type I seesaw model.

The search for sterile neutrinos is an active area of particle physics and cosmology.
The sterile neutrinos may be responsible for a number of unexplained phenomena in particle physics, cosmology and astrophysics.
Sterile neutrinos will be further searched for indirectly and directly in future experiments, which would unravel
the puzzle of SBL neutrino anomalies and make them a very testable particle by providing even more stringent constraints  in particle physics and cosmology.
If the mass of sterile neutrino is smaller than the energies of particles in the experiments, there will be a change to probe them in the laboratory due to the production either through mixing between active and sterile neutrinos or in high energy particle collisions.

%
%

\section*{Acknowledgments}
This work was supported by the Research Program  funded  by the SeoulTech (Seoul National University of Science and Technology).



\begin{thebibliography}{0}    
\def\plb#1#2#3{Phys.\ Lett.\       {\bf B#1}, (#3) #2}
\def\mpl#1#2#3{Mod.\ Phys.\ Lett.\ {\bf A#1}, (#3) #2}
\def\rep#1#2#3{Phys.\ Rep.\        {\bf #1},  (#3) #2}
\def\sci#1#2#3{Science             {\bf #1},  (#3) #2}
\def\astro#1#2#3{Astrophys.\ J.\   {\bf #1},  (#3) #2}
\def\epj#1#2#3{Eur.\ Phys.\ J.  {\bf C#1},  (#3) #2}
\def\jhep#1#2#3{JHEP               {\bf #1},  (#3) #2}
\def\jpg#1#2#3{J.\ Phys.\        {\bf G#1},  (#3) #2}
\def\ijmp#1#2#3{Int.\ J.\ Mod.\ Phys.\ {\bf #1},  (#3) #2}
\def\ptp#1#2#3{Prog.\ Theor.\ Phys.\ {\bf #1},  (#3) #2}
\expandafter\ifx\csname url\endcsname\relax
\def\url#1{{\tt #1}}\fi
\expandafter\ifx\csname urlprefix\endcsname\relax\def\urlprefix{URL }\fi
\providecommand{\eprint}[2][]{\url{#2}}
\bibitem{atm} 
  Y.~Fukuda {\it et al.} [Super-Kamiokande Collaboration],
  Phys.\ Rev.\ Lett.\  {\bf 81}, 1562 (1998).
\bibitem{solar} 
  Q.~R.~Ahmad {\it et al.} [SNO Collaboration],
  Phys.\ Rev.\ Lett.\  {\bf 87}, 071301 (2001).
\bibitem{re1} 
  F.~P.~An {\it et al.} [Daya Bay Collaboration],
  Phys.\ Rev.\ Lett.\  {\bf 108}, 171803 (2012).
\bibitem{re2} 
  J.~K.~Ahn {\it et al.} [RENO Collaboration],
  Phys.\ Rev.\ Lett.\  {\bf 108}, 191802 (2012).
\bibitem{re3} 
  Y.~Abe {\it et al.} [Double Chooz Collaboration],
  Phys.\ Rev.\ D {\bf 86}, 052008 (2012);  Phys.\ Rev.\ Lett.\  {\bf 108}, 131801 (2012).
\bibitem{ALEPH:2005ab} 
  S.~Schael {\it et al.} [ALEPH and DELPHI and L3 and OPAL and SLD Collaborations and LEP Electroweak Working Group and SLD Electroweak Group and SLD Heavy Flavour Group],
  Phys.\ Rept.\  {\bf 427}, 257 (2006).
\bibitem{review}
See for review,
C. Giunti and C.W.Kim {\em Fundamentals of Neutrino Physics and Astrophysics} (Oxford,UK: Oxford UNiversity Press) (2007); Zhi-shong Xing and Shun Zhou {\em Neutrinos in Particle Physics, Astrophysics and Cosmology} (Springer-Verlag Berlin Heidelberg) (2011).
\bibitem{pdg}
M. Tanabashi {\em et al.} (Particle Data Group) Phys. Rev. D {\bf 98}, 030001 (2018).
\bibitem{CPV}
  H.~Nunokawa, S.~J.~Parke and J.~W.~F.~Valle,
  Prog.\ Part.\ Nucl.\ Phys.\  {\bf 60}, 338 (2008);
G.~C.~Branco, R.~G.~Felipe and F.~R.~Joaquim,
  Rev.\ Mod.\ Phys.\  {\bf 84}, 515 (2012);
 S.~F.~King,
  Prog.\ Part.\ Nucl.\ Phys.\  {\bf 94}, 217 (2017);
 S.~T.~Petcov,
  Eur.\ Phys.\ J.\ C {\bf 78}, no. 9, 709 (2018).

\bibitem{Vagnozzi:2017ovm}
  S.~Vagnozzi, E.~Giusarma, O.~Mena, K.~Freese, M.~Gerbino, S.~Ho and M.~Lattanzi,
  Phys.\ Rev.\ D {\bf 96}, no. 12, 123503 (2017).
\bibitem{mass-ordering2}
  R.~B.~Patterson,
  Ann.\ Rev.\ Nucl.\ Part.\ Sci.\  {\bf 65}, 177 (2015).

\bibitem{Giunti:2019aiy} 
For a recent review, see
  C.~Giunti and T.~Lasserre,
  arXiv:1901.08330 [hep-ph].

\bibitem{Pontecorvo:1967fh} 
  B.~Pontecorvo,
  Sov.\ Phys.\ JETP {\bf 26}, 984 (1968).

\bibitem{Volkas:2001zb} 
  R.~R.~Volkas,
  Prog.\ Part.\ Nucl.\ Phys.\  {\bf 48}, 161 (2002).
\bibitem{Mohapatra:2006gs} 
  R.~N.~Mohapatra and A.~Y.~Smirnov,
  Ann.\ Rev.\ Nucl.\ Part.\ Sci.\  {\bf 56}, 569 (2006).
\bibitem{Boyarsky:2009ix} 
  A.~Boyarsky, O.~Ruchayskiy and M.~Shaposhnikov,
  Ann.\ Rev.\ Nucl.\ Part.\ Sci.\  {\bf 59}, 191 (2009).
\bibitem{Giunti:2015wnd} 
  C.~Giunti,
  Nucl.\ Phys.\ B {\bf 908}, 336 (2016).

\bibitem{Gariazzo:2015rra} 
  S.~Gariazzo, C.~Giunti, M.~Laveder, Y.~F.~Li and E.~M.~Zavanin,
  J.\ Phys.\ G {\bf 43}, 033001 (2016).

\bibitem{Adhikari:2016bei} 
  M.~Drewes {\it et al.},
  JCAP {\bf 1701}, no. 01, 025 (2017).
\bibitem{Abazajian:2017tcc} 
  K.~N.~Abazajian,
  Phys.\ Rept.\  {\bf 711-712}, 1 (2017).
\bibitem{Lattanzi:2017ubx} 
  M.~Lattanzi and M.~Gerbino,
  Front.\ in Phys.\  {\bf 5}, 70 (2018).

\bibitem{Boyarsky:2018tvu} 
  A.~Boyarsky, M.~Drewes, T.~Lasserre, S.~Mertens and O.~Ruchayskiy,
  Prog.\ Part.\ Nucl.\ Phys.\  {\bf 104}, 1 (2019).

\bibitem{seesaw}
P.~Minkowski,
  Phys.\ Lett.\  {\bf 67B}, 421 (1977);
M.  Gell-Mann, P. Ramond and R. Slansky, {\rm Supergravity}, ed. by D. Freedman and P. Van Nieuwenhuizen, North Holland, Amsterdam, 315 (1979);
T. Yanagida, Proc. of the {\em Workshop on the Unified Theory and Baryon Number in the Universe}, ed. by O. Sawada and A. Sugamoto (KEK report 79-18), 95 (1979);
  R.~N.~Mohapatra and G.~Senjanovic,
  Phys.\ Rev.\ Lett.\  {\bf 44}, 912 (1980);
 J.~Schechter and J.~W.~F.~Valle,
  Phys.\ Rev.\ D {\bf 22}, 2227 (1980).
\bibitem{Antusch:2016ejd} 
  S.~Antusch, E.~Cazzato and O.~Fischer,
  Int.\ J.\ Mod.\ Phys.\ A {\bf 32}, no. 14, 1750078 (2017).
\bibitem{Mohapatra:1986bd} 
See also,
  R.~N.~Mohapatra and J.~W.~F.~Valle,
  Phys.\ Rev.\ D {\bf 34}, 1642 (1986);
  M.~Malinsky, J.~C.~Romao and J.~W.~F.~Valle,
  Phys.\ Rev.\ Lett.\  {\bf 95}, 161801 (2005);
  M.~Shaposhnikov,
  Nucl.\ Phys.\ B {\bf 763}, 49 (2007);
  J.~Kersten and A.~Y.~Smirnov,
  Phys.\ Rev.\ D {\bf 76}, 073005 (2007);
  S.~Antusch and O.~Fischer,
  JHEP {\bf 1505}, 053 (2015).

\bibitem{Han:2006ip} 
  T.~Han and B.~Zhang,
  Phys.\ Rev.\ Lett.\  {\bf 97}, 171804 (2006).
\bibitem{Atre:2009rg} 
  A.~Atre, T.~Han, S.~Pascoli and B.~Zhang,
  JHEP {\bf 0905}, 030 (2009).
\bibitem{Dev:2013wba} 
  P.~S.~B.~Dev, A.~Pilaftsis and U.~k.~Yang,
  Phys.\ Rev.\ Lett.\  {\bf 112}, no. 8, 081801 (2014).
\bibitem{Alva:2014gxa} 
  D.~Alva, T.~Han and R.~Ruiz,
  JHEP {\bf 1502}, 072 (2015).

\bibitem{Dodelson:1993je} 
  S.~Dodelson and L.~M.~Widrow,
  Phys.\ Rev.\ Lett.\  {\bf 72}, 17 (1994).

\bibitem{leptogenesis}
M.~Fukugita and T.~Yanagida,
  Phys.\ Lett.\ B {\bf 174}, 45 (1986);
W.~Buchmuller, R.~D.~Peccei and T.~Yanagida,
  Ann.\ Rev.\ Nucl.\ Part.\ Sci.\  {\bf 55}, 311 (2005);
S.~Davidson, E.~Nardi and Y.~Nir,
  Phys.\ Rept.\  {\bf 466}, 105 (2008).

\bibitem{Athanassopoulos:1996jb} 
Athanassopoulos C {\it et al.} [LSND Collaboration] {\em Phys. Rev. Lett. {\bf 75}},
2650 (1995) [nucl-ex/9504002];
  C.~Athanassopoulos {\it et al.} [LSND Collaboration],
  {\em Phys.\ Rev.\ Lett.\  {\bf 77}}, 3082 (1996)
  doi:10.1103/PhysRevLett.77.3082
  [nucl-ex/9605003].

\bibitem{Aguilar:2001ty} 
  A.~Aguilar-Arevalo {\it et al.} [LSND Collaboration],
  {\em Phys.\ Rev.\ D {\bf 64}}, 112007 (2001)
  doi:10.1103/PhysRevD.64.112007
  [hep-ex/0104049].

\bibitem{Armbruster:2002mp} 
  B.~Armbruster {\it et al.} [KARMEN Collaboration],
  Phys.\ Rev.\ D {\bf 65}, 112001 (2002)
  doi:10.1103/PhysRevD.65.112001
  [hep-ex/0203021].

\bibitem{AguilarArevalo:2007it} 
  A.~A.~Aguilar-Arevalo {\it et al.} [MiniBooNE Collaboration],
  Phys.\ Rev.\ Lett.\  {\bf 98}, 231801 (2007)
  doi:10.1103/PhysRevLett.98.231801
  [arXiv:0704.1500 [hep-ex]].

\bibitem{AguilarArevalo:2008rc} 
  A.~A.~Aguilar-Arevalo {\it et al.} [MiniBooNE Collaboration],
  Phys.\ Rev.\ Lett.\  {\bf 102}, 101802 (2009)
  doi:10.1103/PhysRevLett.102.101802
  [arXiv:0812.2243 [hep-ex]].

\bibitem{Aguilar-Arevalo:2018gpe} 
  A.~A.~Aguilar-Arevalo {\it et al.} [MiniBooNE Collaboration],
  Phys.\ Rev.\ Lett.\  {\bf 121}, no. 22, 221801 (2018)
  doi:10.1103/PhysRevLett.121.221801
  [arXiv:1805.12028 [hep-ex]].

\bibitem{Antonello:2012pq} 
  M.~Antonello {\it et al.},
  Eur.\ Phys.\ J.\ C {\bf 73}, no. 3, 2345 (2013)
  doi:10.1140/epjc/s10052-013-2345-6
  [arXiv:1209.0122 [hep-ex]].

\bibitem{Farnese:2015kfa} 
  C.~Farnese,
  AIP Conf.\ Proc.\  {\bf 1666}, 110002 (2015).
  doi:10.1063/1.4915574

\bibitem{Agafonova:2013xsk} 
  N.~Agafonova {\it et al.} [OPERA Collaboration],
  JHEP {\bf 1307}, 004 (2013)
  Addendum: [JHEP {\bf 1307}, 085 (2013)]
  doi:10.1007/JHEP07(2013)004, 10.1007/JHEP07(2013)085
  [arXiv:1303.3953 [hep-ex]].

\bibitem{reactorantineutrinoanomaly}
Mention G.et al., Phys. Rev. {\bf D 83}, 073006 (2011).


\bibitem{Kaether:2010ag} 
  F.~Kaether, W.~Hampel, G.~Heusser, J.~Kiko and T.~Kirsten,
  Phys.\ Lett.\ B {\bf 685}, 47 (2010)
  doi:10.1016/j.physletb.2010.01.030
  [arXiv:1001.2731 [hep-ex]].

\bibitem{Abdurashitov:2005tb} 
  J.~N.~Abdurashitov {\it et al.},
  Phys.\ Rev.\ C {\bf 73}, 045805 (2006)
  [nucl-ex/0512041].

\bibitem{Mueller:2011nm} 
  T.~A.~Mueller {\it et al.},
  Phys.\ Rev.\ C {\bf 83}, 054615 (2011)
  [arXiv:1101.2663 [hep-ex]].

\bibitem{Huber:2011wv} 
  P.~Huber,
  Phys.\ Rev.\ C {\bf 84}, 024617 (2011)
  Erratum: [Phys.\ Rev.\ C {\bf 85}, 029901 (2012)]
  [arXiv:1106.0687 [hep-ph]].

\bibitem{previous}

P.~Vogel, G.~K.~Schenter, F.~M.~Mann and R.~E.~Schenter,
  Phys.\ Rev.\ C {\bf 24}, 1543 (1981);
  K.~Schreckenbach, G.~Colvin, W.~Gelletly and F.~Von Feilitzsch,
  Phys.\ Lett.\  {\bf 160B}, 325 (1985);
  A.~A.~Hahn, K.~Schreckenbach, G.~Colvin, B.~Krusche, W.~Gelletly and F.~Von Feilitzsch,
  Phys.\ Lett.\ B {\bf 218}, 365 (1989).

\bibitem{Declais:1995su}
Achkar B, et~al.
\textit{Nucl. Phys.} B434,503 (1995).
\bibitem{Declais:1994ma}
Declais Y, et~al.
\textit{Phys. Lett.} B338, 383 (1994).
\bibitem{Kuvshinnikov:1990ry}
Kuvshinnikov A, et~al.
\textit{JETP Lett.} 54:253 (1991).
\bibitem{Apollonio:2002gd}
Apollonio M, et~al.
\textit{Eur. Phys. J.} C27:331 (2003).
\bibitem{Zacek:1986cu}
Zacek G, et~al.
\textit{Phys. Rev.} D34:2621 (1986).
\bibitem{Kwon:1981ua}
Kwon H, et~al.
\textit{Phys. Rev.} D24:1097 (1981).

\bibitem{Hoummada:1995zz}
Hoummada A, et~al.
\textit{Applied Radiation and Isotopes} 46:449 (1995).
\bibitem{Vidyakin:1994ut}
Vidyakin GS, et~al.
\textit{JETP Lett.} 59:390 (1994).

\bibitem{Boireau:2015dda}
Boireau G, et~al.
\textit{Phys. Rev.} D93:112006 (2016).
\bibitem{Afonin:1988gx}
Afonin AI, et~al.
\textit{Sov. Phys. JETP} 67:213 (1988).
\bibitem{Greenwood:1996pb}
Greenwood ZD, et~al.
\textit{Phys. Rev.} D53:6054 (1996).
\bibitem{Boehm:2001ik}
Boehm F, et~al.
\textit{Phys. Rev.} D64:112001 (2001).

\bibitem{An:2017osx}
An FP, et~al.
\textit{Phys.Rev.Lett.} 118:251801 (2017).

\bibitem{Adey:2018qct}
Adey D, et~al. arXiv:1808.10836 [hep-ex] (2018).
\bibitem{RENO:2018pwo}
Bak G, et~al. arXiv:1806.00574 [hep-ex] (2018).
\bibitem{Abe:2014bwa}
Abe Y, et~al.
\textit{JHEP} 10:086 (2014), [Erratum: JHEP 02, 074 (2015)].

\bibitem{Bezerra-NOW2018}
Bezerra T (2018), talk presented at {NOW 2018, 9-16 September 2018}.


\bibitem{Ko:2016owz} 
  Y.~J.~Ko {\it et al.} [NEOS Collaboration],
  Phys.\ Rev.\ Lett.\  {\bf 118}, no. 12, 121802 (2017).

\bibitem{Alekseev:2018efk} 
  I.~Alekseev {\it et al.} [DANSS Collaboration],
  Phys.\ Lett.\ B {\bf 787}, 56 (2018).

\bibitem{An:2015nua} 
  F.~P.~An {\it et al.} [Daya Bay Collaboration],
  Phys.\ Rev.\ Lett.\  {\bf 116}, no. 6, 061801 (2016)
  Erratum: [Phys.\ Rev.\ Lett.\  {\bf 118}, no. 9, 099902 (2017)].

\bibitem{RENO:2015ksa} 
  J.~H.~Choi {\it et al.} [RENO Collaboration],
  Phys.\ Rev.\ Lett.\  {\bf 116}, no. 21, 211801 (2016).

\bibitem{gallium}
 M.~Laveder,
  Nucl.\ Phys.\ Proc.\ Suppl.\  {\bf 168}, 344 (2007); C.~Giunti and M.~Laveder,
  Mod.\ Phys.\ Lett.\ A {\bf 22}, 2499 (2007);
 C.~Giunti and M.~Laveder,
  Phys.\ Rev.\ C {\bf 83}, 065504 (2011);
 C.~Giunti, M.~Laveder, Y.~F.~Li, Q.~Y.~Liu and H.~W.~Long,
  Phys.\ Rev.\ D {\bf 86}, 113014 (2012).

\bibitem{Bilenky:1996rw} 
  S.~M.~Bilenky, C.~Giunti and W.~Grimus,
  Eur.\ Phys.\ J.\ C {\bf 1}, 247 (1998) see also
 C.~Giunti and M.~Laveder,
  Phys.\ Rev.\ D {\bf 84}, 073008 (2011).

\bibitem{Gariazzo:2017fdh} 
  S.~Gariazzo, C.~Giunti, M.~Laveder and Y.~F.~Li,
  JHEP {\bf 1706}, 135 (2017).

\bibitem{mass-dif}
C. Giunti and C. W. Kim, Fundamentals of Neutrino Physics and Astrophysics (Oxford University
Press, Oxford, UK, 2007) pp. 1–728.

\bibitem{3+1-prob}
   S.~M.~Bilenky, C.~Giunti and W.~Grimus,
  Prog.\ Part.\ Nucl.\ Phys.\  {\bf 43}, 1 (1999)
  doi:10.1016/S0146-6410(99)00092-7
  [hep-ph/9812360].

\bibitem{tension-formular}
N. Okada abd O. Yasuda, In. J. Mod. Phys. A12, 3669 (1997).

\bibitem{Dentler:2018sju} 
  M.~Dentler, Á.~Hernández-Cabezudo, J.~Kopp, P.~A.~N.~Machado, M.~Maltoni, I.~Martinez-Soler and T.~Schwetz,
  JHEP {\bf 1808}, 010 (2018).

\bibitem{Giunti:2016oan} 
  C.~Giunti,
  J.\ Phys.\ Conf.\ Ser.\  {\bf 888}, no. 1, 012019 (2017)
  [J.\ Phys.\ Conf.\ Ser.\  {\bf 888}, no. 1, 012231 (2017)]
  doi:10.1088/1742-6596/888/1/012019, 10.1088/1742-6596/888/1/012231
  [arXiv:1609.04688 [hep-ph]].

\bibitem{Giunti:2015mwa} 
  C.~Giunti and E.~M.~Zavanin,
  Mod.\ Phys.\ Lett.\ A {\bf 31}, no. 01, 1650003 (2015)

\bibitem{Giunti:2011gz}
 J.~Kopp, M.~Maltoni and T.~Schwetz,
  Phys.\ Rev.\ Lett.\  {\bf 107}, 091801 (2011);
  C.~Giunti and M.~Laveder,
  Phys.\ Rev.\ D {\bf 84}, 073008 (2011);
  Phys.\ Rev.\ D {\bf 84}, 093006 (2011);
  Phys.\ Lett.\ B {\bf 706}, 200 (2011);
M.~Archidiacono, N.~Fornengo, C.~Giunti and A.~Melchiorri,
  Phys.\ Rev.\ D {\bf 86}, 065028 (2012);
 J.~M.~Conrad, C.~M.~Ignarra, G.~Karagiorgi, M.~H.~Shaevitz and J.~Spitz,
  Adv.\ High Energy Phys.\  {\bf 2013}, 163897 (2013);
C.~Giunti and E.~M.~Zavanin,
  Mod.\ Phys.\ Lett.\ A {\bf 31}, no. 01, 1650003 (2015).


\bibitem{Adamson:2017uda} 
  P.~Adamson {\it et al.} [MINOS+ Collaboration],
  Phys.\ Rev.\ Lett.\  {\bf 122}, no. 9, 091803 (2019).

\bibitem{TheIceCube:2016oqi} 
  M.~G.~Aartsen {\it et al.} [IceCube Collaboration],
  Phys.\ Rev.\ Lett.\  {\bf 117}, no. 7, 071801 (2016).

\bibitem{Giunti:2013aea}
 C.~Giunti, M.~Laveder, Y.~F.~Li and H.~W.~Long,
  Phys.\ Rev.\ D {\bf 88}, 073008 (2013).

\bibitem{AguilarArevalo:2013pmq}
 A.~A.~Aguilar-Arevalo {\it et al.} [MiniBooNE Collaboration],
  Phys.\ Rev.\ Lett.\  {\bf 110}, 161801 (2013).

\bibitem{Giunti:2011hn} 
  C.~Giunti and M.~Laveder,
  Phys.\ Rev.\ D {\bf 84}, 093006 (2011);
  Phys.\ Lett.\ B {\bf 706}, 200 (2011).
\bibitem{Bilenky:1999wz} 
  S.~M.~Bilenky, C.~Giunti, W.~Grimus, B.~Kayser and S.~T.~Petcov,
  Phys.\ Lett.\ B {\bf 465}, 193 (1999).
\bibitem{Goswami:2005ng} 
  S.~Goswami and W.~Rodejohann,
  Phys.\ Rev.\ D {\bf 73}, 113003 (2006).
\bibitem{Goswami:2007kv} 
  S.~Goswami and W.~Rodejohann,
  JHEP {\bf 0710}, 073 (2007).
\bibitem{Barry:2011wb} 
  J.~Barry, W.~Rodejohann and H.~Zhang,
  JHEP {\bf 1107}, 091 (2011).
\bibitem{Li:2011ss} 
  Y.~F.~Li and S.~s.~Liu,
  Phys.\ Lett.\ B {\bf 706}, 406 (2012).
\bibitem{Girardi:2013zra} 
  I.~Girardi, A.~Meroni and S.~T.~Petcov,
  JHEP {\bf 1311}, 146 (2013).
\bibitem{Giunti:2015kza} 
  C.~Giunti and E.~M.~Zavanin,
  JHEP {\bf 1507}, 171 (2015).
\bibitem{Liu:2017ago} 
  J.~H.~Liu and S.~Zhou,
  Int.\ J.\ Mod.\ Phys.\ A {\bf 33}, no. 02, 1850014 (2018).
\bibitem{Jang:2018zug} 
  C.~H.~Jang, B.~J.~Kim, Y.~J.~Ko and K.~Siyeon,
  J.\ Korean Phys.\ Soc.\  {\bf 73}, no. 11, 1625 (2018).
\bibitem{Schechter:1981bd} 
  J.~Schechter and J.~W.~F.~Valle,
  Phys.\ Rev.\ D {\bf 25}, 2951 (1982).
\bibitem{Duerr:2011zd} 
  M.~Duerr, M.~Lindner and A.~Merle,
  JHEP {\bf 1106}, 091 (2011).
\bibitem{Liu:2016oph} 
  J.~H.~Liu, J.~Zhang and S.~Zhou,
  Phys.\ Lett.\ B {\bf 760}, 571 (2016).
\bibitem{Rodejohann:2011mu} 
For reviews, see
  W.~Rodejohann,
  Int.\ J.\ Mod.\ Phys.\ E {\bf 20}, 1833 (2011);
  S.~M.~Bilenky and C.~Giunti,
  Mod.\ Phys.\ Lett.\ A {\bf 27}, 1230015 (2012);
  W.~Rodejohann,
  J.\ Phys.\ G {\bf 39}, 124008 (2012);
  S.~M.~Bilenky and C.~Giunti,
  Int.\ J.\ Mod.\ Phys.\ A {\bf 30}, no. 04n05, 1530001 (2015);
  H.~Päs and W.~Rodejohann,
  New J.\ Phys.\  {\bf 17}, no. 11, 115010 (2015);
  S.~Dell'Oro, S.~Marcocci, M.~Viel and F.~Vissani,
  Adv.\ High Energy Phys.\  {\bf 2016}, 2162659 (2016).
\bibitem{Pascoli:2002xq} 
  S.~Pascoli and S.~T.~Petcov,
  Phys.\ Lett.\ B {\bf 544}, 239 (2002).
\bibitem{Xing:2016ymd} 
  Z.~z.~Xing and Z.~h.~Zhao,
  Eur.\ Phys.\ J.\ C {\bf 77}, no. 3, 192 (2017).
\bibitem{Benato:2015via} 
  G.~Benato,
  Eur.\ Phys.\ J.\ C {\bf 75}, no. 11, 563 (2015).
\bibitem{Ge:2016tfx} 
  S.~F.~Ge and M.~Lindner,
  Phys.\ Rev.\ D {\bf 95}, no. 3, 033003 (2017).
%
%
\bibitem{Shvartsman:1969mm} 
  V.~F.~Shvartsman,
  Pisma Zh.\ Eksp.\ Teor.\ Fiz.\  {\bf 9}, 315 (1969)
  [JETP Lett.\  {\bf 9}, 184 (1969)].
\bibitem{Steigman:1977kc} 
  G.~Steigman, D.~N.~Schramm and J.~E.~Gunn,
  Phys.\ Lett.\ B {\bf 66}, 202 (1977)
  [Phys.\ Lett.\  {\bf 66B}, 202 (1977)].

\bibitem{Mangano:2005cc} 
  G.~Mangano, G.~Miele, S.~Pastor, T.~Pinto, O.~Pisanti and P.~D.~Serpico,
  Nucl.\ Phys.\ B {\bf 729}, 221 (2005).

\bibitem{Lesgourgues:2012uu} 
  J.~Lesgourgues and S.~Pastor,
  Adv.\ High Energy Phys.\  {\bf 2012}, 608515 (2012).

\bibitem{Acero:2008rh}
  M.~A.~Acero and J.~Lesgourgues,
  Phys.\ Rev.\ D {\bf 79}, 045026 (2009).

\bibitem{Ade:2013zuv} 
  P.~A.~R.~Ade {\it et al.} [Planck Collaboration],
  Astron.\ Astrophys.\  {\bf 571}, A16 (2014).

\bibitem{Ade:2015xua} 
  P.~A.~R.~Ade {\it et al.} [Planck Collaboration],
  Astron.\ Astrophys.\  {\bf 594}, A13 (2016).


\bibitem{Abazajian:2014gza} 
  K.~N.~Abazajian,
  Phys.\ Rev.\ Lett.\  {\bf 112}, no. 16, 161303 (2014).


\bibitem{Steigman:2012ve}
 G.~Steigman,
  Adv.\ High Energy Phys.\  {\bf 2012}, 268321 (2012).

\bibitem{Iocco:2008va}
F.~Iocco, G.~Mangano, G.~Miele, O.~Pisanti and P.~D.~Serpico,
  Phys.\ Rept.\  {\bf 472}, 1 (2009).

\bibitem{Jacques:2013xr}
T.~D.~Jacques, L.~M.~Krauss and C.~Lunardini,
  Phys.\ Rev.\ D {\bf 87}, no. 8, 083515 (2013)
  Erratum: [Phys.\ Rev.\ D {\bf 88}, no. 10, 109901 (2013)].

\bibitem{Fields:2014uja}
 B.~D.~Fields, P.~Molaro and S.~Sarkar,
  Chin.\ Phys.\ C {\bf 38}, 339 (2014).

\bibitem{Mangano:2011ar}
 G.~Mangano and P.~D.~Serpico,
  Phys.\ Lett.\ B {\bf 701}, 296 (2011).

\bibitem{Lesgourgues:2014zoa} 
  J.~Lesgourgues and S.~Pastor,
  New J.\ Phys.\  {\bf 16}, 065002 (2014).


\bibitem{Cyburt:2015mya}
R.~H.~Cyburt, B.~D.~Fields, K.~A.~Olive and T.~H.~Yeh,
  Rev.\ Mod.\ Phys.\  {\bf 88}, 015004 (2016).

\bibitem{Anderson:2013zyy}
L.~Anderson {\it et al.} [BOSS Collaboration],
  Mon.\ Not.\ Roy.\ Astron.\ Soc.\  {\bf 441}, no. 1, 24 (2014).

\bibitem{Ross:2014qpa}
  A.~J.~Ross, L.~Samushia, C.~Howlett, W.~J.~Percival, A.~Burden and M.~Manera,
  Mon.\ Not.\ Roy.\ Astron.\ Soc.\  {\bf 449}, no. 1, 835 (2015).

\bibitem{Beutler:2011hx}
  F.~Beutler {\it et al.},
  Mon.\ Not.\ Roy.\ Astron.\ Soc.\  {\bf 416}, 3017 (2011).

\bibitem{Feng:2017nss} 
  L.~Feng, J.~F.~Zhang and X.~Zhang,
  Eur.\ Phys.\ J.\ C {\bf 77}, no. 6, 418 (2017).

\bibitem{Cuesta:2015mqa} 
  A.~J.~Cuesta {\it et al.},
  Mon.\ Not.\ Roy.\ Astron.\ Soc.\  {\bf 457}, no. 2, 1770 (2016)

\bibitem{Riess:2016jrr} 
  A.~G.~Riess {\it et al.},
  Astrophys.\ J.\  {\bf 826}, no. 1, 56 (2016)

\bibitem{Ade:2015fva} 
  P.~A.~R.~Ade {\it et al.} [Planck Collaboration],
  Astron.\ Astrophys.\  {\bf 594}, A24 (2016).

\bibitem{Ade:2015zua} 
  P.~A.~R.~Ade {\it et al.} [Planck Collaboration],
  Astron.\ Astrophys.\  {\bf 594}, A15 (2016).


\bibitem{Heymans:2013fya} 
  C.~Heymans {\it et al.},
  Mon.\ Not.\ Roy.\ Astron.\ Soc.\  {\bf 432}, 2433 (2013).


\bibitem{Dolgov:2000ew} 
  A.~D.~Dolgov and S.~H.~Hansen,
  Astropart.\ Phys.\  {\bf 16}, 339 (2002).

\bibitem{Langacker:1989sv} 
  P.~Langacker,
  UPR-0401T.

\bibitem{Abazajian:2001nj} 
  K.~Abazajian, G.~M.~Fuller and M.~Patel,
  Phys.\ Rev.\ D {\bf 64}, 023501 (2001).

\bibitem{Merle:2015vzu} 
  A.~Merle, A.~Schneider and M.~Totzauer,
  JCAP {\bf 1604}, no. 04, 003 (2016).

\bibitem{Fryer:2005sz} 
  C.~L.~Fryer and A.~Kusenko,
  Astrophys.\ J.\ Suppl.\  {\bf 163}, 335 (2006).

\bibitem{Kusenko:1997sp} 
  A.~Kusenko and G.~Segre,
  Phys.\ Lett.\ B {\bf 396}, 197 (1997).

\bibitem{Fuller:2003gy} 
  G.~M.~Fuller, A.~Kusenko, I.~Mocioiu and S.~Pascoli,
  Phys.\ Rev.\ D {\bf 68}, 103002 (2003).

\bibitem{Barkovich:2004jp} 
  M.~Barkovich, J.~C.~D'Olivo and R.~Montemayor,
  Phys.\ Rev.\ D {\bf 70}, 043005 (2004).

\bibitem{Barbieri:1989ti} 
  R.~Barbieri and A.~Dolgov,
  Phys.\ Lett.\ B {\bf 237}, 440 (1990).

\bibitem{MSW}
Mikheyev and S. O. Smirnov, Sov. J. Nucl. Phys., {\bf 42}, 913 (1985);
L. Wolfenstein, Phys. Rev. {\bf D 17}, 2369 (1978).

\bibitem{Enqvist:1990ek} 
  K.~Enqvist, K.~Kainulainen and J.~Maalampi,
  Phys.\ Lett.\ B {\bf 249}, 531 (1990).
\bibitem{Shi:1998km} 
  X.~D.~Shi and G.~M.~Fuller,
  Phys.\ Rev.\ Lett.\  {\bf 82}, 2832 (1999).
\bibitem{Ghiglieri:2015jua} 
  J.~Ghiglieri and M.~Laine,
  JHEP {\bf 1511}, 171 (2015).

\bibitem{Kishimoto:2008ic} 
  C.~T.~Kishimoto and G.~M.~Fuller,
  Phys.\ Rev.\ D {\bf 78}, 023524 (2008).

\bibitem{Laine:2008pg} 
  M.~Laine and M.~Shaposhnikov,
  JCAP {\bf 0806}, 031 (2008).

\bibitem{Bulbul:2014sua} 
  E.~Bulbul, M.~Markevitch, A.~Foster, R.~K.~Smith, M.~Loewenstein and S.~W.~Randall,
  Astrophys.\ J.\  {\bf 789}, 13 (2014).

\bibitem{Boyarsky:2014jta} 
  A.~Boyarsky, O.~Ruchayskiy, D.~Iakubovskyi and J.~Franse,
  Phys.\ Rev.\ Lett.\  {\bf 113}, 251301 (2014)

\bibitem{Schneider:2016uqi} 
  A.~Schneider,
  JCAP {\bf 1604}, no. 04, 059 (2016).

\bibitem{Merle:2017jfn} 
  A.~Merle,
  PoS NOW {\bf 2016}, 082 (2017).

\bibitem{Shaposhnikov:2006xi} 
  M.~Shaposhnikov and I.~Tkachev,
  Phys.\ Lett.\ B {\bf 639}, 414 (2006).
\bibitem{Bezrukov:2009yw} 
  F.~Bezrukov and D.~Gorbunov,
  JHEP {\bf 1005}, 010 (2010).
\bibitem{Kusenko:2006rh} 
  A.~Kusenko,
  Phys.\ Rev.\ Lett.\  {\bf 97}, 241301 (2006).
\bibitem{Petraki:2007gq} 
  K.~Petraki and A.~Kusenko,
  Phys.\ Rev.\ D {\bf 77}, 065014 (2008).

\bibitem{Merle:2013wta} 
  A.~Merle, V.~Niro and D.~Schmidt,
  JCAP {\bf 1403}, 028 (2014).
\bibitem{Boyanovsky:2008nc} 
  D.~Boyanovsky,
  Phys.\ Rev.\ D {\bf 78}, 103505 (2008).
\bibitem{Roland:2014vba} 
  S.~B.~Roland, B.~Shakya and J.~D.~Wells,
  Phys.\ Rev.\ D {\bf 92}, no. 11, 113009 (2015).
\bibitem{Matsui:2015maa} 
  H.~Matsui and M.~Nojiri,
  Phys.\ Rev.\ D {\bf 92}, no. 2, 025045 (2015).
\bibitem{Roland:2016gli} 
  S.~B.~Roland and B.~Shakya,
  JCAP {\bf 1705}, no. 05, 027 (2017).
\bibitem{Frigerio:2014ifa} 
  M.~Frigerio and C.~E.~Yaguna,
  Eur.\ Phys.\ J.\ C {\bf 75}, no. 1, 31 (2015).
\bibitem{Kang:2014mea} 
  S.~K.~Kang and A.~Patra,
  J.\ Korean Phys.\ Soc.\  {\bf 69}, no. 8, 1375 (2016).
\bibitem{Drewes:2015eoa} 
  M.~Drewes and J.~U.~Kang,
  JHEP {\bf 1605}, 051 (2016).
\bibitem{Adulpravitchai:2015mna} 
  A.~Adulpravitchai and M.~A.~Schmidt,
  JHEP {\bf 1512}, 023 (2015).
\bibitem{Shakya:2016oxf} 
  B.~Shakya and J.~D.~Wells,
  Phys.\ Rev.\ D {\bf 96}, no. 3, 031702 (2017).
\bibitem{Shuve:2014doa} 
  B.~Shuve and I.~Yavin,
  Phys.\ Rev.\ D {\bf 89}, no. 11, 113004 (2014).
\bibitem{Abada:2014zra} 
  A.~Abada, G.~Arcadi and M.~Lucente,
  JCAP {\bf 1410}, 001 (2014).

\bibitem{dec-FIMP}
  A.~Merle and M.~Totzauer,
  JCAP {\bf 1506} 011 (2015);
  J.~K\"onig, A.~Merle, and M.~Totzauer,
  JCAP {\bf 1611} (2016) 038.
\bibitem{Pal:1981rm} 
  P.~B.~Pal and L.~Wolfenstein,
  Phys.\ Rev.\ D {\bf 25}, 766 (1982).
\bibitem{Barger:1995ty} 
  V.~D.~Barger, R.~J.~N.~Phillips and S.~Sarkar,
  Phys.\ Lett.\ B {\bf 352}, 365 (1995).
\bibitem{Lee:1977tib} 
  B.~W.~Lee and R.~E.~Shrock,
  Phys.\ Rev.\ D {\bf 16}, 1444 (1977).
\bibitem{Shrock:1982sc} 
  R.~E.~Shrock,
  Nucl.\ Phys.\ B {\bf 206}, 359 (1982).

\bibitem{Abazajian:2001vt} 
  K.~Abazajian, G.~M.~Fuller and W.~H.~Tucker,
  Astrophys.\ J.\  {\bf 562}, 593 (2001).

\bibitem{Mapelli:2005hq} 
  M.~Mapelli and A.~Ferrara,
  Mon.\ Not.\ Roy.\ Astron.\ Soc.\  {\bf 364}, 2 (2005).

\bibitem{Boyarsky:2005us} 
  A.~Boyarsky, A.~Neronov, O.~Ruchayskiy and M.~Shaposhnikov,
  Mon.\ Not.\ Roy.\ Astron.\ Soc.\  {\bf 370}, 213 (2006).

\bibitem{Abazajian:2006jc} 
  K.~N.~Abazajian, M.~Markevitch, S.~M.~Koushiappas and R.~C.~Hickox,
  Phys.\ Rev.\ D {\bf 75}, 063511 (2007).
\bibitem{Boyarsky:2006fg} 
  A.~Boyarsky, A.~Neronov, O.~Ruchayskiy, M.~Shaposhnikov and I.~Tkachev,
  Phys.\ Rev.\ Lett.\  {\bf 97}, 261302 (2006).
\bibitem{Mateo:1998wg} 
  M.~Mateo,
  Ann.\ Rev.\ Astron.\ Astrophys.\  {\bf 36}, 435 (1998).
\bibitem{Boyarsky:2009rb} 
  A.~Boyarsky, O.~Ruchayskiy, D.~Iakubovskyi, A.~V.~Maccio' and D.~Malyshev,
  arXiv:0911.1774 [astro-ph.CO].
\bibitem{Boyarsky:2009af} 
  A.~Boyarsky, A.~Neronov, O.~Ruchayskiy and I.~Tkachev,
  Phys.\ Rev.\ Lett.\  {\bf 104}, 191301 (2010).

\bibitem{Wyman:2013lza} 
  M.~Wyman, D.~H.~Rudd, R.~A.~Vanderveld and W.~Hu,
  Phys.\ Rev.\ Lett.\  {\bf 112}, no. 5, 051302 (2014).
\bibitem{Hamann:2013iba} 
  J.~Hamann and J.~Hasenkamp,
  JCAP {\bf 1310}, 044 (2013).
\bibitem{Battye:2013xqa} 
  R.~A.~Battye and A.~Moss,
  Phys.\ Rev.\ Lett.\  {\bf 112}, no. 5, 051303 (2014).
\bibitem{Leistedt:2014sia} 
  B.~Leistedt, H.~V.~Peiris and L.~Verde,
  Phys.\ Rev.\ Lett.\  {\bf 113}, 041301 (2014).
\bibitem{Palanque-Delabrouille:2014jca} 
  N.~Palanque-Delabrouille {\it et al.},
  JCAP {\bf 1502}, no. 02, 045 (2015).
\bibitem{Borde:2014xsa} 
  A.~Borde, N.~Palanque-Delabrouille, G.~Rossi, M.~Viel, J.~S.~Bolton, C.~Yèche, J.~M.~LeGoff and J.~Rich,
  JCAP {\bf 1407}, 005 (2014).
\bibitem{York:2000gk} 
  D.~G.~York {\it et al.} [SDSS Collaboration],
  Astron.\ J.\  {\bf 120}, 1579 (2000).

\bibitem{McDonald:2004eu} 
  P.~McDonald {\it et al.} [SDSS Collaboration],
  Astrophys.\ J.\ Suppl.\  {\bf 163}, 80 (2006)
\bibitem{McDonald:2004xn} 
  P.~McDonald {\it et al.} [SDSS Collaboration],
  Astrophys.\ J.\  {\bf 635}, 761 (2005)

\bibitem{Eisenstein:2011sa} 
  D.~J.~Eisenstein {\it et al.} [SDSS Collaboration],
  Astron.\ J.\  {\bf 142}, 72 (2011).
\bibitem{Dawson:2012va} 
  K.~S.~Dawson {\it et al.} [BOSS Collaboration],
  Astron.\ J.\  {\bf 145}, 10 (2013).

\bibitem{Markovic:2013iza} 
  K.~Markovič and M.~Viel,
  Publ.\ Astron.\ Soc.\ Austral.\  {\bf 31}, e006 (2014).

\bibitem{Bond:1980ha} 
  J.~R.~Bond, G.~Efstathiou and J.~Silk,
  Phys.\ Rev.\ Lett.\  {\bf 45}, 1980 (1980).
\bibitem{Viel:2013apy} 
  M.~Viel, G.~D.~Becker, J.~S.~Bolton and M.~G.~Haehnelt,
  Phys.\ Rev.\ D {\bf 88}, 043502 (2013).
\bibitem{Bhattacharyya:2017epv} 
  S.~Bhattacharyya, H.~Motz, S.~Torii and Y.~Asaoka,
  JCAP {\bf 1708}, no. 08, 012 (2017).
\bibitem{Irsic:2017ixq} 
  V.~Iršič {\it et al.},
  Phys.\ Rev.\ D {\bf 96}, no. 2, 023522 (2017).

\bibitem{Kang:2006sn} 
  S.~K.~Kang and C.~S.~Kim,
  Phys.\ Lett.\ B {\bf 646}, 248 (2007).

\bibitem{Covi:1996wh} 
  L.~Covi, E.~Roulet and F.~Vissani,
  Phys.\ Lett.\ B {\bf 384}, 169 (1996).
\bibitem{SungCheon:2007nw} 
  H.~Sung Cheon, S.~K.~Kang and C.~S.~Kim,
  JCAP {\bf 0805}, 004 (2008).
\bibitem{Krasnikov:1997nh} 
  N.~V.~Krasnikov,
  Mod.\ Phys.\ Lett.\ A {\bf 13}, 893 (1998).
\bibitem{Babu:2014pxa} 
  K.~S.~Babu and R.~N.~Mohapatra,
  Phys.\ Rev.\ D {\bf 89}, 115011 (2014).
\bibitem{Merle:2015oja} 
  A.~Merle and M.~Totzauer,
  JCAP {\bf 1506}, 011 (2015).
%

\bibitem{Viel:2005qj} 
  M.~Viel, J.~Lesgourgues, M.~G.~Haehnelt, S.~Matarrese and A.~Riotto,
  Phys.\ Rev.\ D {\bf 71}, 063534 (2005).
\bibitem{Seljak:2006qw} 
  U.~Seljak, A.~Makarov, P.~McDonald and H.~Trac,
  Phys.\ Rev.\ Lett.\  {\bf 97}, 191303 (2006).
\bibitem{Viel:2006kd} 
  M.~Viel, J.~Lesgourgues, M.~G.~Haehnelt, S.~Matarrese and A.~Riotto,
  Phys.\ Rev.\ Lett.\  {\bf 97}, 071301 (2006).
\bibitem{McDonald:1993ex} 
  J.~McDonald,
  Phys.\ Rev.\ D {\bf 50}, 3637 (1994).


\bibitem{Buchmuller:2002rq} 
  W.~Buchmuller, P.~Di Bari and M.~Plumacher,
  Nucl.\ Phys.\ B {\bf 643}, 367 (2002).
\bibitem{Pilaftsis:1997jf} 
  A.~Pilaftsis,
  Phys.\ Rev.\ D {\bf 56}, 5431 (1997);
Nucl. Phys. B{\bf 504}, 61 (1997);  A.~Pilaftsis and T. E. J. Underwood, Nucl.Phys., B {\bf 692}, 303 (2004).





\end{thebibliography}
\end{document}